\documentclass[%
 letterpaper,twocolumn,english,
 superscriptaddress,
showpacs,preprintnumbers,
 amsmath,amssymb,
 aps,
 prc,
floatfix,
]{revtex4-1}

\usepackage[usenames]{color}
\usepackage{amsmath}
\usepackage{amssymb}
\usepackage{longtable}  
\usepackage{graphicx}
\usepackage[dvipdfm,ps2pdf,unicode=true,bookmarks=false,breaklinks,pdfborder={0
    0 0},backref=false,colorlinks=true]{hyperref}  
\hypersetup{pdftitle={First measurement of unpolarized SIDIS cross section and cross section ratios from a $^3$He target},pdfauthor={Xuefei Yan , et. al. (The Jefferson Lab Hall
    A Collaboration)},pdfsubject={v1: nucl-ex,
    hep-ex},pdfkeywords={Transverse DSA Inclusive hadron SIDIS He-3
    $A_\text{LT}$ TMD},linkcolor=DarkBlueCite, citecolor=DarkBlueCite,
  urlcolor=DarkBlueCite}  
\usepackage[hyphenbreaks]{breakurl}

\makeatletter

\usepackage{dcolumn}
\usepackage{bm}

\definecolor{DarkBlueCite}{rgb}{0.1,0.0,0.5}

\makeatother

\begin{document}



\author{X.~Yan}\email[Corresponding author: ]{xy33@phy.duke.edu}
\affiliation{Duke University, Durham, NC 27708}
\author{K.~Allada}
\affiliation{Massachusetts Institute of Technology, Cambridge, MA 02139}
\affiliation{Thomas Jefferson National Accelerator Facility, Newport
  News, VA 23606}  
\author{K.~Aniol}
\affiliation{California State University, Los Angeles, Los Angeles, CA 90032}
\author{J.R.M.~Annand}
\affiliation{University of Glasgow, Glasgow G12 8QQ, Scotland, United Kingdom}
\author{T.~Averett}
\affiliation{College of William and Mary, Williamsburg, VA 23187}
\author{F.~Benmokhtar}
\affiliation{Duquesne University, Pittsburgh, PA 15282}
\author{W.~Bertozzi}
\affiliation{Massachusetts Institute of Technology, Cambridge, MA 02139}
\author{P.C.~Bradshaw}
\affiliation{College of William and Mary, Williamsburg, VA 23187}
\author{P.~Bosted}
\affiliation{Thomas Jefferson National Accelerator Facility, Newport
  News, VA 23606}
\author{A.~Camsonne}
\affiliation{Thomas Jefferson National Accelerator Facility, Newport News, VA 23606}
% ...
\collaboration{The Jefferson Lab Hall A Collaboration}
\noaffiliation

\title{First measurement of unpolarized SIDIS cross section from a $^3$He target} 


\begin{abstract}
The unpolarized semi-inclusive deep-inelastic scattering (SIDIS) differential cross sections in $^3$He($e,e^{\prime}\pi^{\pm}$)$X$ have been measured for the first time in Jefferson Lab experiment E06-010 performed with a $5.9\,$GeV $e^-$  beam on a $^3$He target. The experiment focuses on the valence quark region, covering a kinematic range $0.12 < x_{bj} < 0.45$, $1 < Q^2 < 4 \, \textrm{(GeV/c)}^2$, $0.45 < z_{h} < 0.65$, and $0.05 < P_t < 0.55 \, \textrm{GeV/c}$.  The extracted SIDIS differential cross sections of $\pi^{\pm}$ production are compared with existing phenomenological models while the $^3$He nucleus approximated as two protons and one neutron in a plane wave picture, in multi-dimensional bins. Within the experimental uncertainties, the azimuthal modulations of the cross sections are found to be consistent with zero.

\end{abstract}

\pacs{}

\maketitle


\section{\label{intro} Introduction}

One of the main goals in nuclear and particle physics is to unravel ultimately the nucleon structure in terms of quarks and gluons, the fundamental degrees of freedom of quantum chromodynamics (QCD). Due to the nonperturbative nature of QCD at hadronic scales, it is not possible yet to calculate the structures of hadrons directly from first principles of QCD. The lepton-nucleon/nucleus deep inelastic scattering is an important experimental approach and has been widely employed for more than 40 years. During the last decade or so, both experimental and theoretical studies have revealed the nontrivial effects of quark intrinsic transverse momentum, especially spin-related, probed by the SIDIS processes.

In polarized and unpolarized SIDIS processes, azimuthal modulations of cross sections were found to be sizable \cite{EMC_mod,HERMES_0,COMPASS_0,CLAS_0}. The intrinsic transverse momenta of the quarks are expected to play an important role in the observed modulations \cite{Anselmino_2005,Anselmino_0}. To incorporate the intrinsic transverse momentum carried by the partons in the description of the SIDIS processes, Transverse Momentum Dependent (TMD) Parton Distribution Functions (PDFs) and Fragmentation Functions (FFs) were proposed \cite{Mulders_TMD,BM_original}. TMD PDFs and FFs include dependence on the transverse momentum of the partons in addition to the longitudinal momentum used in the traditional one-dimensional PDFs and FFs, and can provide more complete understanding of the nucleon structure. A TMD factorization formalism  was developed, incorporating the TMD PDFs and FFs \cite{Collins_tmd_fac,Brodsky_tmd_fac,Ji_tmd_fac,Aybat_review}. Within the TMD factorization framework, plus additional simplifications and assumptions, the 18 structure functions comprising the SIDIS differential cross section, are expressed as the convolutions of TMD PDFs and FFs \cite{Bacchetta_basis} (naive $x$-$z$ factorization). TMD PDFs and FFs have been parameterized and utilized in the phenomenological studies of the world data of SIDIS and $e^+e^-$ annihilation \cite{Barone_review_2010,Barone2015, Anselmino2014, Bacchetta2011}. An example showing the power of this factorization scheme is the agreement between the model description and the experiment for the Sivers and Collins effects \cite{Barone_review_2010}. The Sivers effect emerges from the convolution of the Sivers TMD PDF and the unpolarized TMD FF. The Collins effect is from the convolution of the transversity TMD PDF and the Collins TMD FF. Sivers and Collins effects are related to different azimuthal modulations in SIDIS differential cross sections with transversely polarized nucleon \cite{Barone_review_2010,Bacchetta_basis}. Nontrivial azimuthal modulations in unpolarized SIDIS processes arise from the convolution of unpolarized TMD PDF and FF with factors involving the quark intrinsic transverse momentum, known as the Cahn effect \cite{Cahn_original}, and the convolution of the Boer-Mulders function and the Collins function, known as the Boer-Mulders effect \cite{BM_original}. Various TMD PDFs provide valuable anatomy of the nucleon structure. For instance, the Boer-Mulders TMD PDF describes the distribution of transversely polarized quarks inside an unpolarized nucleon \cite{Barone_review_2010}.

While factorization originates in the high energy limit ($Q \gg \Lambda_{\rm{QCD}}$ or $Q \gg M_{\rm{nucleon}}$) \cite{Collins_book, PDG}, and at low $Q^2$ the description using hadronic degrees of freedom are more widely used \cite{crisis}, the applicability of the quark-parton model with factorization in modest $Q^2$ ranges has been observed in quark-hadron duality \cite{HallC_PRC,CLAS_PRD}. One needs to note that at modest $Q^2$ ranges, higher-twist terms suppressed by powers of $(1/Q)$ would be larger than those in the range of large $Q^2$, and could bring non-negligible effects \cite{Barone2015}.

While SIDIS measurements on the proton have been carried out by a number of experiments \cite{Barone_review_2010,Barone2015, Anselmino2014, Bacchetta2011,HallC_PRC,CLAS_PRD,HERMES_1,COMPASS_1,HERMES_2,COMPASS_2} and more data will be available, SIDIS data on the neutron are rather limited. Since there is no stable neutron target, using a polarized $^3$He target as an effective polarized neutron target for experimental studies related to the spin structure of the neutron is uniquely advantageous, due to the dominant neutron spin contribution to the $^3$He spin~\cite{3He_spin}. The SIDIS experiment E06-010 in Hall A of Jefferson Lab was carried out with a 5.9 GeV polarized electron beam and a transversely polarized $^3$He target, between October 2008 and February 2009. The experiment covered a kinematic range $0.12 < x_{bj} < 0.45$, $1 < Q^2 < 4 \; (\textrm{GeV/c})^2$, $0.45 < z_{h} < 0.65$, and $0.05 < P_t < 0.55 \; \textrm{GeV/c}$. Studies on the data of E06-010 for single-spin asymmetries (SSAs) and double-spin asymmetries (DSAs) have been carried out \cite{XQ,JH,YZ,YXZ}. These first SIDIS asymmetry results from $^3$He as an effective neutron target were related to TMD PDFs such as transversity, Sivers, pretzelosity, trans-helicity (g$_{1T} ^q$) and TMD FFs such as Collins.

The unpolarized SIDIS differential cross section, while the spin dependent azimuthal modulations are canceled, still involves nontrivial modulations from the Cahn and Boer-Mulders effects. The unpolarized SIDIS differential cross section in the quark-parton model as well as the parameterization of the related TMD PDFs and FFs are presented in section \ref{sec-theory}. As in the studies of the world data \cite{Anselmino_2005,Barone2015, Anselmino2014, Bacchetta2011}, the SIDIS cross section is expressed in the functional form based on the quark-parton model with naive $x$-$z$ factorization, and the transverse momentum dependence described as a Gaussian distribution. In global analyses fitting different types of data (multiplicities and/or asymmetries) in different kinematic ranges, very different values were extracted for the width of the quark intrinsic transverse momentum, $\langle k_{\perp}^2 \rangle$. Namely $\langle k_{\perp}^2 \rangle$ is at the level of 0.2 GeV$^2$ in \cite{Anselmino_2005,Bacchetta2011}, at the level of 0.5 GeV$^2$ in \cite{Anselmino2014} and less than 0.05 GeV$^2$ in \cite{Barone2015}. While the multiplicities and asymmetries from experiments have been fitted with ratios of combinations of the theoretical cross sections, as in the studies of the world data \cite{Anselmino_2005,Barone2015, Anselmino2014, Bacchetta2011}, the corresponding study for the absolute cross sections is rather limited.

In addition to the fact that the absolute cross sections provide more complete information than multiplicities and asymmetries (ratios of combinations of the polarized and unpolarized cross sections), TMD evolution also has much stronger effect on the absolute cross sections \cite{Anselmino_phm}. In recent years, the unpolarized SIDIS processes have attracted considerable interest due to providing special insights into the TMD evolution effect \cite{Martin_2016}.

In this paper, using the E06-010 experimental data, we present the first extraction of the unpolarized SIDIS differential cross sections from a $^3$He target, comparisons with different models, the study of azimuthal modulations in the extracted cross sections, and the constraints on the phenomenological parameters from the data in this study. In this paper, the units GeV/c and GeV are not discriminated for conciseness of expressions.

\section{\label{sec-theory} Quark-parton model and SIDIS parameterization}
The processes of interest are the unpolarized SIDIS $e(l)+N(P) \to e^{\prime}(l^{\prime}) +\pi^{\pm}(P_h)+X(P_X)$, where the variables in the parentheses are the four-vector momenta, $e$ the beam electron, $N$ the target nucleon, $e^{\prime}$ the scattered electron being detected, $\pi^{\pm}$ the detected hadron (charged pion) and $X$ the final state particles not being detected. The unpolarized SIDIS differential cross section is expressed as
\begin{widetext}
\begin{eqnarray} \label{SIDIS_xs_def}
\frac{d\sigma}{dx_{bj}dydz_{h} d\phi_S dP_t^2d\phi_h} &=& \frac{\alpha^2}{2Q^2 x_{bj} y} \left [ A\cdot F_{UU}+B\cdot F_{UU}^{\cos\phi_h} \cos\phi_h+C\cdot F_{UU}^{\cos2\phi_h} \cos2\phi_h \right ] ,
\end{eqnarray}
\end{widetext}
 where $\alpha$ is the electromagnetic fine-structure constant, $A=(1+(1-y)^2)$, $B=2(2-y)\sqrt{1-y}$, $C=2(1-y)$, $x_{bj}=Q^2/(2P\cdot q)$, $y=(P\cdot q)/(P\cdot l)$, $z_{h}=(P\cdot P_h)/(P\cdot q)$, $q=l-l^{\prime}$ and $Q^2=-q^2$ \cite{Bacchetta_basis,Barone2015, Anselmino2014, Bacchetta2011}. $\phi_S$ is the azimuthal angle of the nucleon spin direction, and can be integrated out in the unpolarized SIDIS process yielding an additional $2\pi$ factor for the $F_{UU}$'s. The reference frame and the definition of the azimuthal angle $\phi_h$ between the lepton scattering plane and the hadron plane follow the ``Trento Conventions" as in \cite{trento_convention}. The transverse momentum of the detected hadron is denoted as $P_t$.

The structure function $F_{UU}$ involves a convolution of the unpolarized TMD PDF $f_q(x_{bj},k_{\perp})$ and TMD FF $D_q(z_{h},p_{\perp})$, where $k_{\perp}$ is the intrinsic transverse momentum of the parton and $p_{\perp}$ the transverse momentum of the fragmenting hadron with respect to the parton.
The structure function $F_{UU}^{\cos\phi_h}$ at the lowest twist (twist-3), consists of a Cahn contribution and a Boer-Mulders contribution. The structure function $F_{UU}^{\cos2\phi_h}$ consists of a twist-2 Boer-Mulders contribution and a twist-4 Cahn contribution. The Cahn contributions involve the convolution of the unpolarized TMD PDF $f_q(x_{bj},k_{\perp})$ and TMD FF $D_q(z_{h},p_{\perp})$. The Boer-Mulders contributions involve the convolution of the Boer-Mulders TMD PDF $\Delta f_{q\uparrow}(x_{bj},k_{\perp})=-h_1^{\perp}(x_{bj},k_{\perp})\cdot k_{\perp}/M_p$ and the Collins TMD FF $\Delta D_{q\uparrow}(z_{h},p_{\perp})=2p_{\perp}\cdot H_1^{\perp}(z_{h},p_{\perp})/(z_{h}M_h)$. A unit vector is defined for convenience as $\mathbf{h \equiv P_t/|P_t|}$.
The structure functions are given below with the momentum conservation condition $\mathbf{P_t} = z_{h}\mathbf{k_{\perp}}+\mathbf{p_{\perp}}$. 
\begin{widetext}
\begin{eqnarray} \label{FUU}
F_{UU} &=& \sum_q e_q^2 x \int d^2 \mathbf{k_{\perp}} f_q(x_{bj},k_{\perp})D_q(z_{h},p_{\perp}), \\
F_{UU}^{\cos\phi_h}|_{\rm{Cahn}} &=& -2\sum_q e_q^2 x \int d^2 \mathbf{k_{\perp}} \frac{\mathbf{k_{\perp}\cdot h}}{Q} f_q(x_{bj},k_{\perp}) D_q(z_{h},p_{\perp}), \\
F_{UU}^{\cos\phi_h}|_{\rm{BM}} &=& \sum_q e_q^2 x \int d^2 \mathbf{k_{\perp}} \frac{k_{\perp}}{Q} \frac{P_t-z_{h}\mathbf{k_{\perp}\cdot h}}{p_{\perp}} \Delta f_{q\uparrow}(x_{bj},k_{\perp}) \Delta D_{q\uparrow}(z_{h},p_{\perp}), \\
F_{UU}^{\cos2\phi_h}|_{\rm{BM}} &=& \sum_q e_q^2 x \int d^2 \mathbf{k_{\perp}} \frac{P_t \mathbf{k_{\perp}\cdot h}+z_{h}[k_{\perp}^2-2(\mathbf{k_{\perp}\cdot h})^2]}{2k_{\perp}p_{\perp}} \Delta f_{q\uparrow}(x_{bj},k_{\perp}) \Delta D_{q\uparrow}(z_{h},p_{\perp}), \\
F_{UU}^{\cos2\phi_h}|_{\rm{Cahn}} &=& 2\sum_q e_q^2 x \int d^2 \mathbf{k_{\perp}} \frac{2(\mathbf{k_{\perp}\cdot h})^2-k_{\perp}^2}{Q^2} f_q(x_{bj},k_{\perp}) D_q(z_{h},p_{\perp}).
\end{eqnarray}
\end{widetext}
Phenomenologically, the Gaussian ansatz is often utilized in TMD parameterizations. The unpolarized TMD PDF $f_q(x_{bj},k_{\perp})$ and unpolarized TMD FF $D_q(z_{h},p_{\perp})$ are expressed as
\begin{eqnarray} \label{untmdpdf_para}
f_q(x_{bj},k_{\perp}) &=& f_q^c(x_{bj}) e^{-k_{\perp}^2/ \langle k_{\perp}^2 \rangle}/(\pi \langle k_{\perp}^2 \rangle), \\
D_q(z_{h},p_{\perp}) &=& D_q^c(z_{h})  e^{-p_{\perp}^2/ \langle p_{\perp}^2 \rangle}/(\pi \langle p_{\perp}^2 \rangle),
\end{eqnarray}
where $f_q^c(x_{bj})$ is the collinear PDF, $D_q^c(z_{h})$ the collinear FF, $\langle k_{\perp}^2 \rangle$ and $\langle p_{\perp}^2 \rangle$ the Gaussian widths as phenomenological parameters. In addition, the widths $\langle k_{\perp}^2 \rangle$ and $\langle p_{\perp}^2 \rangle$ in different studies have different forms of kinematical dependence: $x_{bj}$ dependence for $\langle k_{\perp}^2 \rangle$ and/or $z_{h}$ dependence for $\langle p_{\perp}^2 \rangle$ \cite{Barone2015, Anselmino2014, Bacchetta2011}. The knowledge about the flavor dependence of $\langle k_{\perp}^2 \rangle$ and $\langle p_{\perp}^2 \rangle$ is limited \cite{TMD_width_flavor}, and flavor independence has been assumed in most of the studies. The Boer-Mulders TMD PDF and Collins TMD FF are parameterized as in \cite{Barone2015}.



\section{Experiment}

The experiment E06-010, as introduced in section \ref{intro} and published studies of this experiment \cite{XQ,JH,YZ,YXZ}, produced data sets with a polarized 5.9$\,$GeV electron beam and a transversely polarized $^3$He gas target. The scattered electrons were recorded by the BigBite spectrometer and the electroproduced pions ($\pi^{\pm}$) were recorded by the High Resolution Spectrometer (HRS). To study the unpolarized SIDIS processes, the data with opposite polarization states were combined. The charge difference between the two opposite beam polarizations for the entire experiment was less than 10ppm \cite{YXZ}. The net $^3$He polarization after data combination is less than 0.5$\%$.

In the experiment, the target system consisted of a 40-cm-long glass cell containing about 10 atm of $^3$He polarized by spin-exchange optical pumping \cite{seop}. The direction of the $^3$He polarization was flipped every 20 minutes. At each flip, the percentage of the $^3$He polarization was measured and recorded. The temperature and density of the $^3$He gas in the target cell was monitored and recorded in the data together with the information from the detectors.

The BigBite spectrometer consisted of a single dipole magnet, eighteen planes of multiwire drift chambers in three groups and a scintillator plane between the lead-glass preshower and shower calorimeters. The knowledge of the magnetic field and the information from the drift chambers, were used to reconstruct the tracks of charged particles. The trigger was formed by summing the signals from the preshower and shower calorimeters. The preshower and shower energy deposition with the reconstructed momentum were utilized for the particle identification (PID) in BigBite \cite{xq_thesis}.

The HRS was configured for hadron detection. The trigger was provided by two scintillator planes. Four detectors in the HRS were used for PID: a CO$_2$ gas $\check{\rm{C}}$erenkov detector for electron identification, an aerogel $\check{\rm{C}}$erenkov detector for pion identification, a ring imaging $\check{\rm{C}}$erenkov (RICH) detector for $\pi^{\pm}$, $K^{\pm}$, and proton identification, and two layers of lead-glass calorimeter for electron-hadron separation \cite{KA, xq_thesis, HRS_NIM}.

\section{Data Analysis}

The data analysis for the unpolarized SIDIS differential cross section is more complicated than that for the asymmetry studies due to the need of a thorough understanding and description of the experimental acceptance as well as a good control of the systematic uncertainties, some of which were less important due to the cancellation in the asymmetry studies. Dedicated developments and updates of the detector models in the simulation enabled a good description of the experimental acceptance, and have been successfully used in single electron channels as well as coincidence SIDIS channels. Detailed studies of the systematic uncertainties have been carried out thoroughly for the cross section extraction and the overall systematic uncertainty is mostly under 10$\%$. In addition, radiative corrections, exclusive tail subtractions and bin-centering corrections have been applied. The additional studies important for the analysis of the SIDIS cross section beyond the asymmetry studies of this experiment \cite{XQ,JH,KA,YZ,YXZ} are presented below.

\subsection{Monte Carlo simulations}

For a full description of the experimental acceptance of E06-010, a model for the BigBite spectrometer used in E06-010 for electron detection has been developed and incorporated into the SIMC package \cite{SIMC_pack} which was initially developed for Hall C experiments and used for the semi-inclusive studies in Hall C \cite{HallC_PRC}. It was adapted for this experiment \cite{SIMC_trans}. It contains a realistic description of various detectors including the HRS used in the experiment E06-010 for hadron detection. The energy loss, multi-scattering, pion decay processes have also been included in the SIMC package.

\begin{figure}[ht]
\begin{center}
\includegraphics[width=0.45\textwidth]{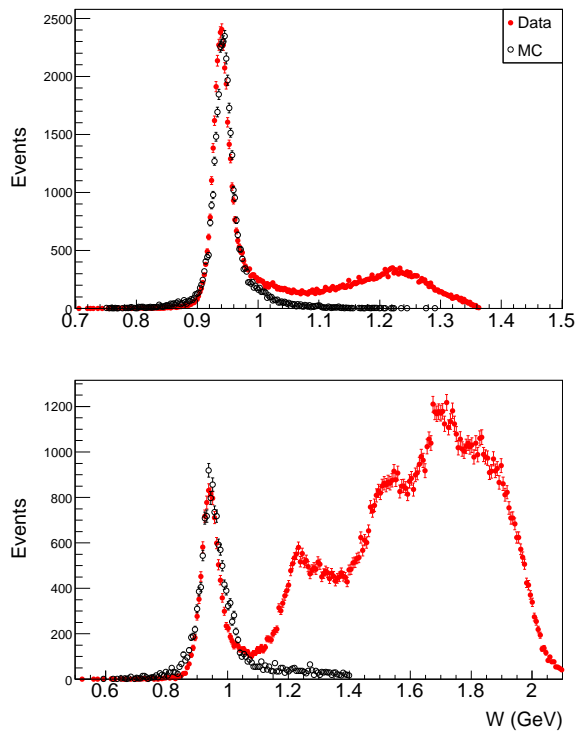}
\caption{(color online). Invariant mass ($W$) bin comparison for 1.23$\,$GeV (top panel) and 2.4$\,$GeV (bottom panel) beam elastic EP calibration run using BigBite. The error bars represent the statistical uncertainties. The red solid circles are from the data. The black open circles are from the simulation. The simulation only covers the elastic EP scattering channel and compares well with the proton $W$ peak while the peaks from higher resonances only appear in the data.}
\label{W_both_plot}
\end{center}
\end{figure}

\begin{figure}[ht]
\begin{center}
\includegraphics[width=0.45\textwidth]{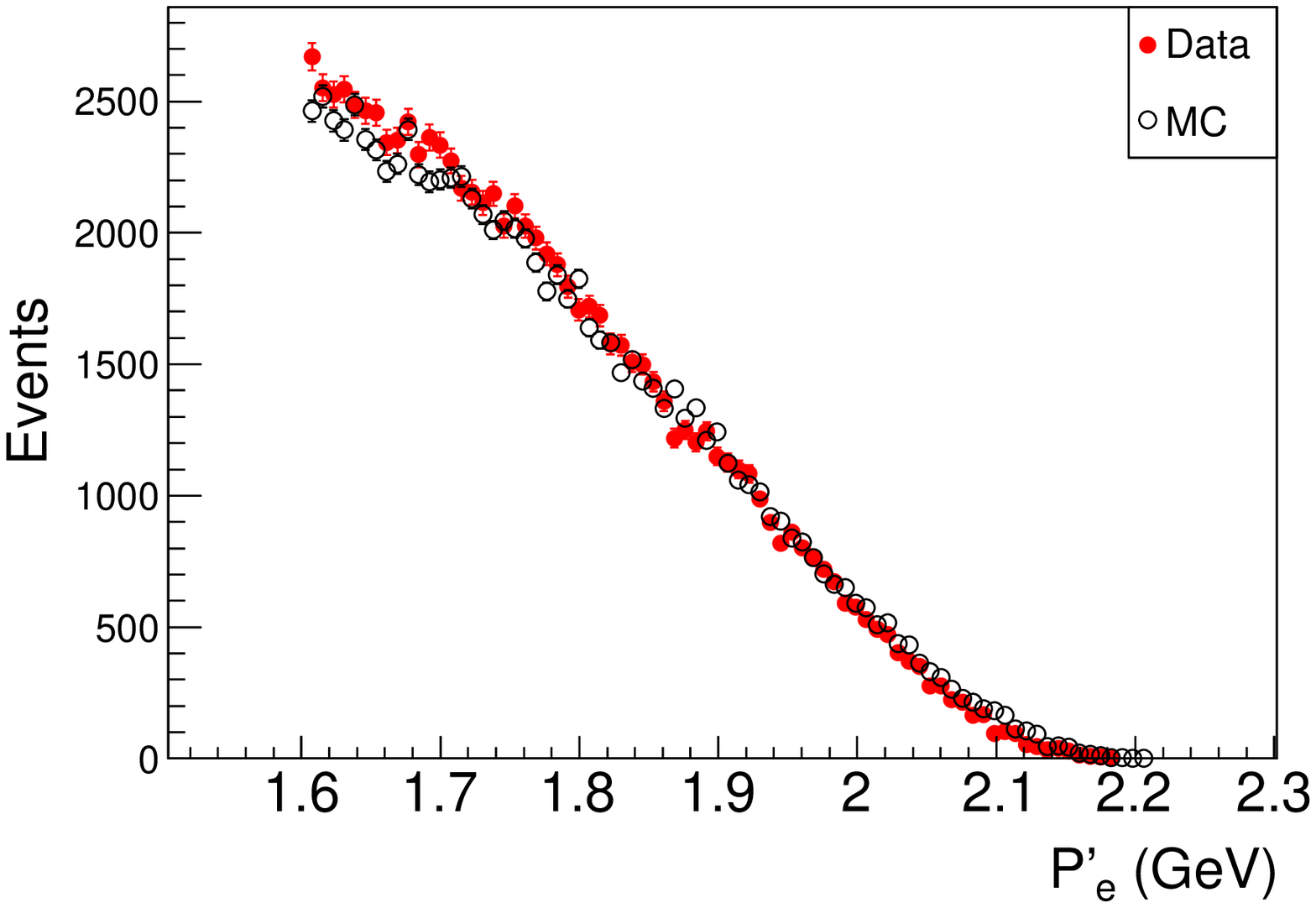}
\caption{(color online). Scattered electron momentum ($P_e^{\prime}$) bin comparison of $^3$He inclusive DIS channel in BigBite. The error bars represent statistical uncertainties. The red solid circles are from the data. The black open circles are from the simulation.}
\label{Inclusive_DIS_BB_plot}
\end{center}
\end{figure}

\begin{figure}[ht]
\begin{center}
\includegraphics[width=0.45\textwidth]{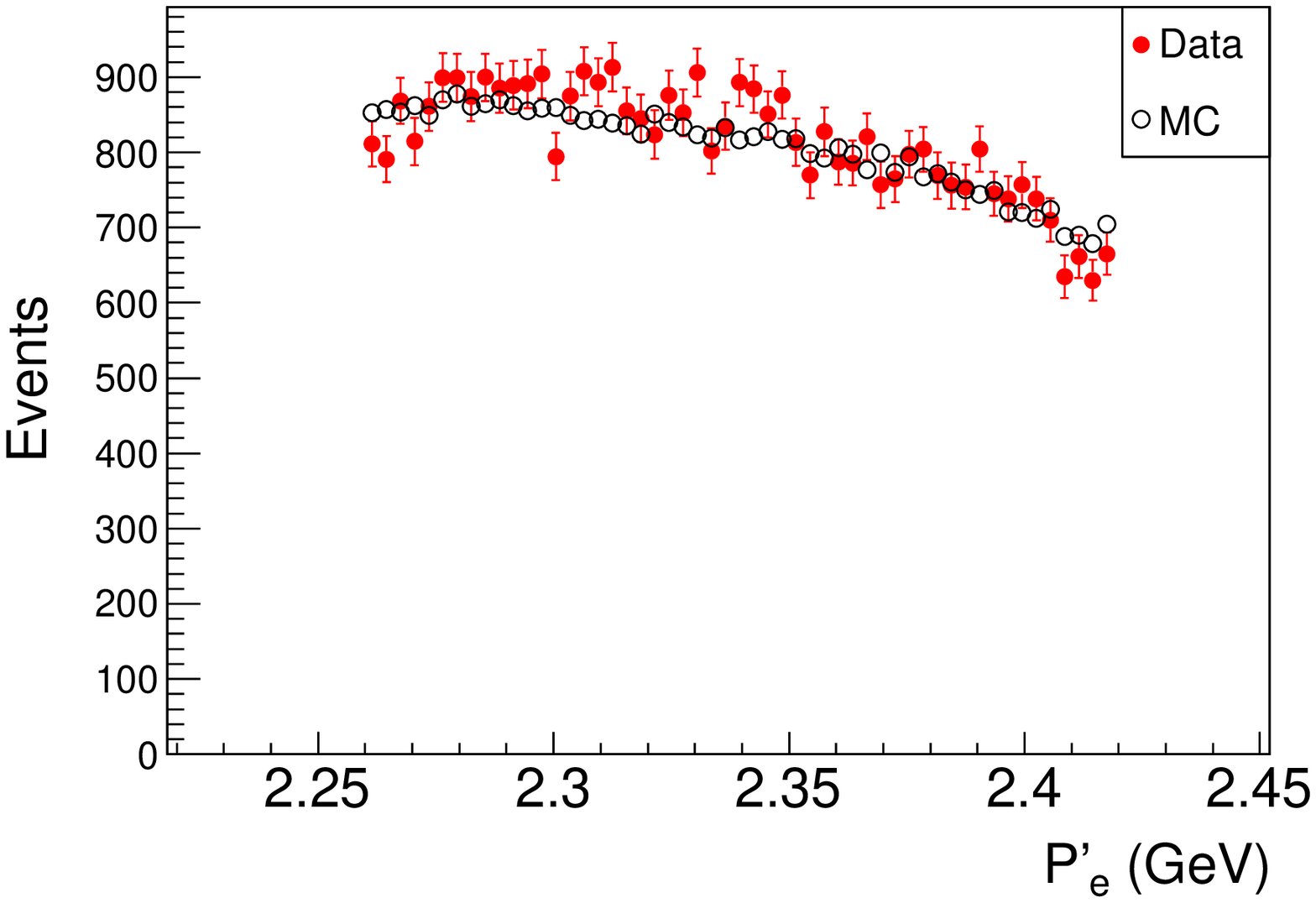}
\caption{(color online). Scattered electron momentum ($P_e^{\prime}$) bin comparison of H$_2$ inclusive DIS channel in the HRS. The error bars represent statistical uncertainties. The red solid circles are from the data. The black open circles are from the simulation.}
\label{Inclusive_H2DIS_HRS_plot}
\end{center}
\end{figure}

\begin{figure}[ht]
\begin{center}
\includegraphics[width=0.45\textwidth]{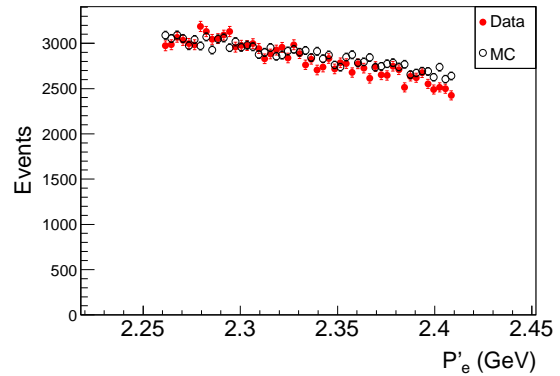}
\caption{(color online). Scattered electron momentum ($P_e^{\prime}$) bin comparison of $^3$He inclusive DIS channel in the HRS. The error bars represent statistical uncertainties. The red solid circles are from the data. The black open circles are from the simulation.}
\label{Inclusive_3HeDIS_HRS_plot}
\end{center}
\end{figure}

The detector model of BigBite was tested by using the calibration runs of elastic electron-proton (EP) scattering at incident electron beam energies of 1.23 and 2.4 GeV (Fig. \ref{W_both_plot}), as well as the inclusive DIS channel from the $^3$He production data at 5.9 GeV (Fig. \ref{Inclusive_DIS_BB_plot}) by using the single BigBite trigger. Inclusive DIS data from H$_2$ reference runs and $^3$He production runs at 5.9 GeV with the single HRS trigger have been used to test the description of the HRS experimental acceptance (Figs. \ref{Inclusive_H2DIS_HRS_plot} and \ref{Inclusive_3HeDIS_HRS_plot}).

\subsection{Data corrections and cross section extraction}

Corrections need to be applied to the raw data for cross section extraction, namely the efficiency correction, contamination/background subtraction,  acceptance corrections, radiative corrections and bin-centering corrections.

The PID of electrons in BigBite was based on the combination of a cut in the preshower energy deposition and a 2D cut in the ratio of total shower energy deposition and the reconstructed momentum in order to suppress the $\pi^-$ contamination. The PID cuts were optimized by minimizing the loss of electron events and maximizing the suppression of the $\pi^-$ contamination. The portions of the remaining $\pi^-$ contamination and the loss of electron events were estimated based on fitting and discriminating the $\pi^-$ and electron signals in the preshower calorimeter. The data after the PID cuts were corrected by subtracting the remaining $\pi^-$ contamination and compensating for the loss of electron events. The remaining $\pi^-$ contamination contributed from less than 0.1$\%$ to 6$\%$ to the signal after the PID cuts, with descending range of electron momentum. The loss of electron events varied from 2$\%$ to 30$\%$ with descending range of electron momentum.

The $\pi^0$, from the electroproduction, decays into two photons. The high energy photons create the photon-induced electron contamination through the pair-production process. The percentage of this contamination in the total electron events was determined by comparing the positron yield in the BigBite with the reversed magnetic field and the electron yield in the BigBite in the production runs. The photon-induced electron contamination was then subtracted. The photon-induced electron contamination contributed 2$\%$ in high electron momentum range and up to 40$\%$ in low electron momentum range, in the total electron events of the BigBite spectrometer.

The PID of $\pi^{\pm}$ in the HRS was based on the combination of gas $\check{\rm{C}}$erenkov, aerogel $\check{\rm{C}}$erenkov and lead glass calorimeter signals. The loss of $\pi^{\pm}$ events ($\sim$5$\%$), negatively charged non-pion hadron and electron contamination to $\pi^-$ events ($<$0.5$\%$) and positively charged non-pion hadron contamination to $\pi^+$ events ($<$1$\%$) have been evaluated and included in the correction.

The experimental acceptance was described by the Monte Carlo simulation with the BigBite and HRS models, both of which have been checked by single electron channels. The acceptance correction was based on the Monte Carlo simulation with the same kinematic cuts as applied to the data. The number of events in each bin from the simulation weighted by the theoretical cross section $N_{sim}$ divided by the non-weighted simulation with the same kinematic cuts in the same bin $N_{phs}$, becomes the (averaged) theoretical cross section in this bin. The same was done for the data events $N_{data}$, forming a quantity $N_{data}/N_{phs}$ in each bin from which the experimental cross section can be determined after applying other data corrections in addition. More details for the acceptance correction can be found in Appendix \ref{acc-corr}.

There were complicated and time-dependent drifts of the total shower energy threshold for the BigBite trigger during the experiment. To address this issue, a total shower energy cut of $E_{tot}> 900$ MeV was used, high enough to override the fluctuations of threshold-related inefficiency, but not too high to significantly reduce the kinematic range and valuable data. A description of the total shower energy deposition was developed and included in the BigBite model of simulation, based on the experimental data from the BigBite calibration runs and checked by the production runs. Simulations with the same total shower energy cut were used for the BigBite trigger efficiency correction.

The tracking efficiency of BigBite was evaluated using the elastic EP scattering with well-known cross sections and state-of-the-art radiative corrections. Two beam energies, 1.23 and 2.4 GeV, were used in the elastic EP runs, covering low and high momentum acceptance of the BigBite spectrometer. The tracking efficiency was also checked using the inclusive DIS channel in the $^3$He production data. The tracking efficiency was estimated to be 73$\%$ to 75$\%$.

The contribution from the exclusive channels $e+p \to e^{\prime}+\pi^++n$ and $e+n \to e^{\prime}+\pi^-+p$ were subtracted using simulation with a cross section model tested in the kinematic range of this experiment \cite{HallC_PRC}. The contributions from the exclusive channels were from 2$\%$ to 7.5$\%$ in the $\pi^+$ production channel and 0.5$\%$ to 3$\%$ in the $\pi^-$ production channel. The internal radiative corrections for the SIDIS channels were carried out using the HAPRAD package \cite{HAPRAD}. The radiative corrections based on Mo and Tsai \cite{MoTsai} built in SIMC were used as a comparison and to estimate systematic uncertainty \cite{HallC_PRC}. The energy loss, multi-scattering effects and $\pi^{\pm}$ decay were simulated using the established components of SIMC, while the radiation length and materials were defined based on the configuration of the experiment E06-010.

The bin-centering corrections were based on the quark-parton model used in comparison with the data. In each of the bins, the central value of each variable involved in the SIDIS cross section was found experimentally and differed by less than 0.5$\%$ from the Monte Carlo simulation. The bin-centering correction for the cross section from the data in one bin is defined as
\begin{eqnarray} \label{eqBCC}
\sigma_{exp}^{BCC}=\frac{\sigma_{exp}^{bin}}{\sigma_{MC}^{bin}}\cdot \sigma_{theory} ,
\end{eqnarray}
where $\sigma_{exp}^{BCC}$ is the SIDIS differential cross section extracted experimentally, after the bin-centering correction with experimental central values of variables $(x_{bj},z_{h},Q^2,\phi_h,P_t)$. $\sigma_{theory}$ is the theoretical SIDIS differential cross section with the same values of variables. $\sigma_{exp}^{bin}$ and $\sigma_{MC}^{bin}$ are the SIDIS differential cross sections in the bin with the corresponding central values, determined from the data and simulation, respectively. $\sigma_{exp}^{bin}$ was determined by applying all the corrections to the raw data, namely the efficiency corrections, contamination/background subtractions, radiative corrections and acceptance corrections. $\sigma_{MC}^{bin}$ was determined by applying the acceptance corrections to the Monte Carlo simulation weighted by the theoretical SIDIS differential cross section.

\subsection{Systematic uncertainties}

The systematic uncertainties related to the electron detection in BigBite were dominated by the determination of the photon-induced electron contamination and the evaluation of the efficiency of the total shower energy cut which was applied to remove the effect from the drift of the threshold. The photon-induced electron contamination contained the uncertainty of determining the positron yield from which a large (up to 50$\%$) portion of $\pi^+$ was evaluated and subtracted. The efficiency of the total shower energy cut involved increasing uncertainties due to increasing effect of the threshold drift, in descending momentum range.

To evaluate the systematic uncertainties above, the cuts on preshower energy deposition, total shower energy deposition and reconstructed momentum have been varied around the optimized values. Each set of cuts corresponds to a specific efficiency and contamination. The data have been corrected under sets of varied cuts besides the optimized one. The root mean square value of the differences of the corrected data yields with varied cut sets from the optimized set have been used to define the systematic uncertainties.

The total systematic uncertainties related to the electron detection in the BigBite are in the range from 3$\%$ to 10$\%$ depending on the kinematics.

The systematic uncertainties of the PID of $\pi^{\pm}$ events in the HRS were determined to be less than 2$\%$, using the well-established techniques in the previous studies of this experiment \cite{XQ,JH,YZ,YXZ,KA,xq_thesis}.

In the coincidence channel for SIDIS, the systematic uncertainties in the experimental acceptance corrections by the simulation were determined by putting a series of kinematic cuts besides the central optimized set to both the data and the simulation. The total systematic uncertainties from the acceptance corrections are between 5$\%$ to 10$\%$ depending on the kinematics.

The systematic uncertainties related to the exclusive tail subtractions and the SIDIS radiative corrections have been evaluated in the same manner as \cite{HallC_PRC}. Specifically, different models of the exclusive channels and the difference between the HAPRAD and the SIMC for the radiative corrections have been used to define the systematic uncertainties. The systematic uncertainties for these items are between 2$\%$ to 6$\%$ depending on the kinematics.

The systematic uncertainties of the results related to the central-value uncertainties of the variables $(x_{bj},z_{h},Q^2,\phi_h,P_t)$ have been evaluated by inserting the variable uncertainties to the bin-centering corrections, thus reflected in the extracted cross sections. The systematic uncertainties related to the bin-centering corrections are less than 3$\%$ with a kinematic dependence.

The main contributions of the systematic uncertainties are listed in Table. \ref{sys_table}.

\begin{table}[hbt]
\caption{Systematic uncertainties.}
\begin{tabular}{l c c}\hline
Source & Range ($\%$) \\ \hline
$e^-$ identification in BigBite & 2.0-8.0 \\
$e^-$ tracking efficiency in BigBite & $<$3.0 \\
$\pi^{\pm}$ identification in the HRS & $<$2.0 \\
Experimental acceptance corrections & 5.0-10.0 \\
Radiative corrections & 1.0-3.5 \\
Exclusive tail subtractions & 1.0-3.0 \\
Bin-centering corrections & $<$3.0 \\ \hline
\end{tabular}
\label{sys_table}
\end{table}

\subsection{Kinematical correlations and binning}

\begin{figure*}
\begin{center}
\includegraphics[width=0.9\textwidth]{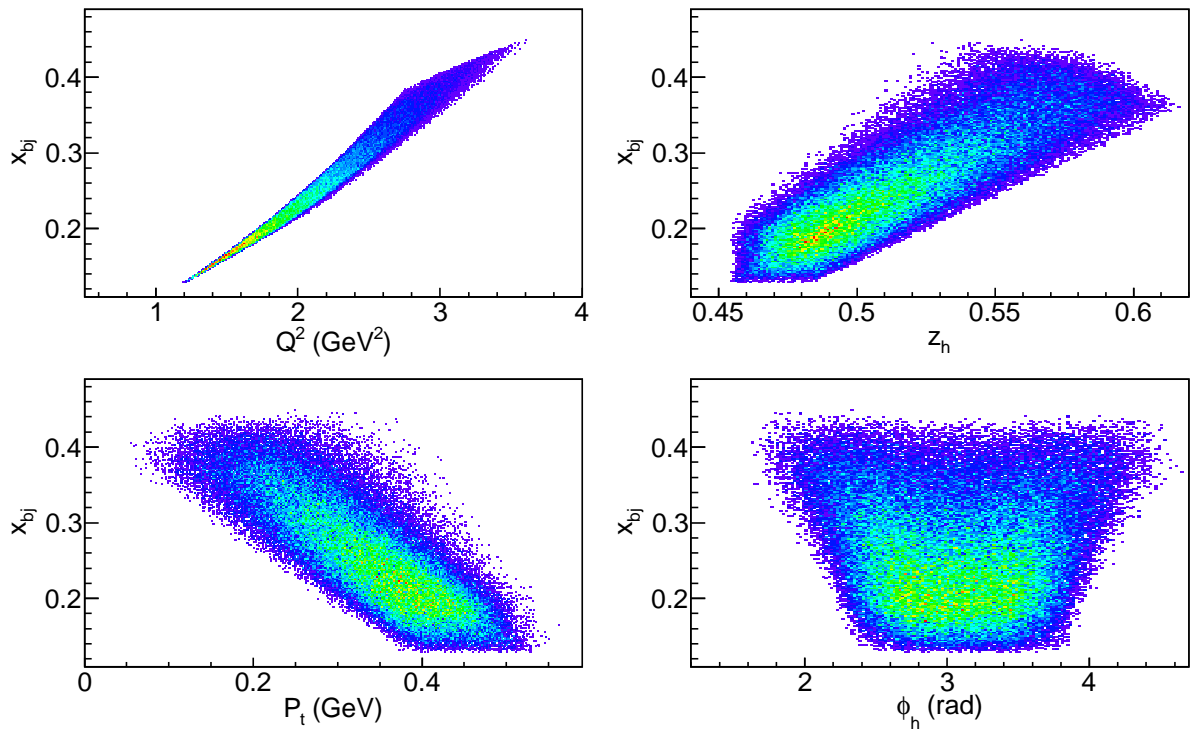}
\caption{(color online). The correlations between $x_{bj}$ and other kinematic variables in experiment E06-010.}
\label{plot_kin_corr}
\end{center}
\end{figure*}

In the production run of E06-010, only one experimental configuration was used. Kinematical correlations are shown in Fig. \ref{plot_kin_corr}. Due to the Kinematical correlations, strict one-dimensional (1D) binning in which only one variable changes while all the other variables stay intact is prohibited.

In this paper, a set of pseudo-one-dimensional (pseudo-1D) bins is used for presenting the results. Pseudo-1D means that when the binning is in one variable, for example, $x_{bj}$, the difference between one bin and another is not only in $x_{bj}$, but in all the variables $(x_{bj},z_{h},Q^2,\phi_h,P_t)$ due to kinematical correlations. Pseudo-1D bins in $x_{bj}$ has 10 consecutive bins with almost equal statistics.

A set of two-dimensional (2D) bins is used to present the $P_t$ dependence of the cross sections. The set of 2D bins (10 $\times$ 10) consists of 10 $P_t$ bins in 10 ranges of $x_{bj}$. The boundaries of the bins are set to make each bin contain almost equal statistics.

A set of three-dimensional (3D) bins is used to present the $\phi_h$ dependence of the cross sections. The data are binned into two ranges of $P_t$ first. In each of the $P_t$ ranges, five $x_{bj}$ bins are defined. 10 $\phi_h$ bins are defined in each of the 2 $\times$ 5 ranges of $P_t$ vs. $x_{bj}$. Each bin of the 2 $\times$ 5 $\times$ 10 set has almost equal statistics.

\section{Results}

The extracted unpolarized SIDIS differential cross sections and the cross section ratios are compared with models in different bin sets in the following sections. Fitting the extracted cross section of the data in this study demonstrates the data's constraint on the parameters describing the SIDIS process. The plane wave impulse approximation (PWIA) treatment of the $^3$He nucleus in the SIDIS process is adopted in this study, thus the modeled SIDIS cross section from $^3$He is the same as the sum of the modeled SIDIS cross sections from two protons and one neutron.

\subsection{\label{xs-1d} Cross sections in pseudo-1D bins}

\begin{figure}
\begin{center}
\includegraphics[width=0.45\textwidth]{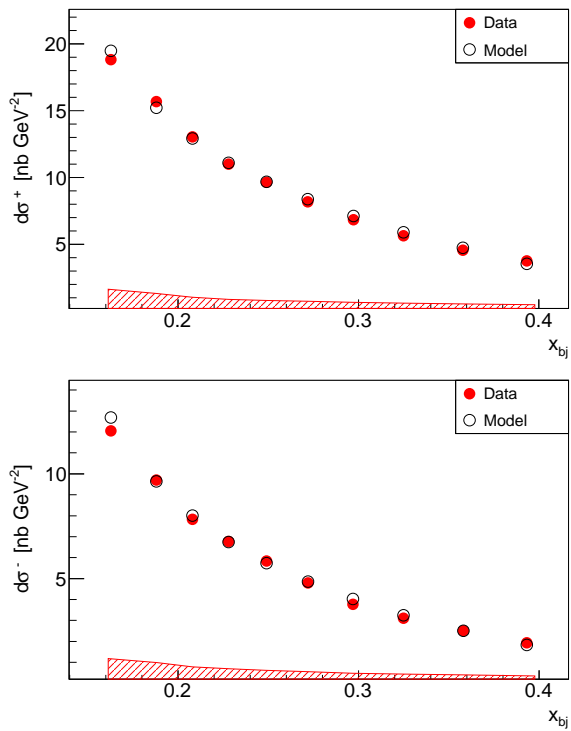}
\caption{(color online). SIDIS differential cross section (defined in text) comparison in pseudo-1D $x_{bj}$ bins. The red solid circles are from the data and the black open circles are from the quark-parton model. The error bar of each point represents the statistical uncertainty, mostly smaller than the markers. The error band on the bottom of each panel represents the experimental systematic uncertainty. The top and bottom panels are for $\pi^+$ and $\pi^-$ production channel, respectively.}
\label{result_xs_x}
\end{center}
\end{figure}

The comparisons of the SIDIS differential cross sections from the data and the quark-parton model in pseudo-1D $x_{bj}$ bins are shown in Fig. \ref{result_xs_x}. The top panel in the figure is for the $\pi^+$ production channel $^3\rm{He}$($e,e^{\prime}\pi^{+}$)$X$  and the bottom panel for the $\pi^-$ production channel $^3\rm{He}$($e,e^{\prime}\pi^{-}$)$X$. The vertical axis is the SIDIS differential cross section $d\sigma/(dx_{bj}dydz_{h} d\phi_S dP_t^2d\phi_h)$ in unit of $\rm{nb}\cdot \rm{GeV}^{-2}\cdot rad^{-2}$. The total experimental systematic uncertainties using quadrature combination of all the sources are shown in the band at the bottom of each plot.

The SIDIS differential cross section from the model is defined in Eq (\ref{SIDIS_xs_def}) and the parameterizations of the Gaussian widths of unpolarized TMD PDF and FF are in the forms as in \cite{Bacchetta2011}, namely $\langle k_{\perp}^2 \rangle =0.14 $ GeV$^2$ and $\langle p_{\perp}^2 \rangle=a \cdot z_{h}^{0.54}(1-z_{h})^{b}$ GeV$^2$, where $a=1.55$ and $b=2.2$ are slightly tuned from the values in one set of the HERMES data analysis inherited and cited by \cite{Bacchetta2011}. The Boer-Mulders TMD PDF and Collins TMD FF parameterizations were taken from \cite{Barone2015}. The effect of the Boer-Mulders terms in the total SIDIS cross sections were found to be less than 2$\%$ in magnitude and opposite in sign for the $\pi^{\pm}$ electroproduction channels. Terms with twists higher than those included in section \ref{sec-theory} were neglected. The model calculates the sum of the cross sections from two protons and one neutron as an approximation for the $^3$He nucleus.

Agreement between the data and the model is observed.

\subsection{The $\phi_h$ dependence of the cross sections}
\begin{figure*}
\begin{center}
\includegraphics[width=0.9\textwidth]{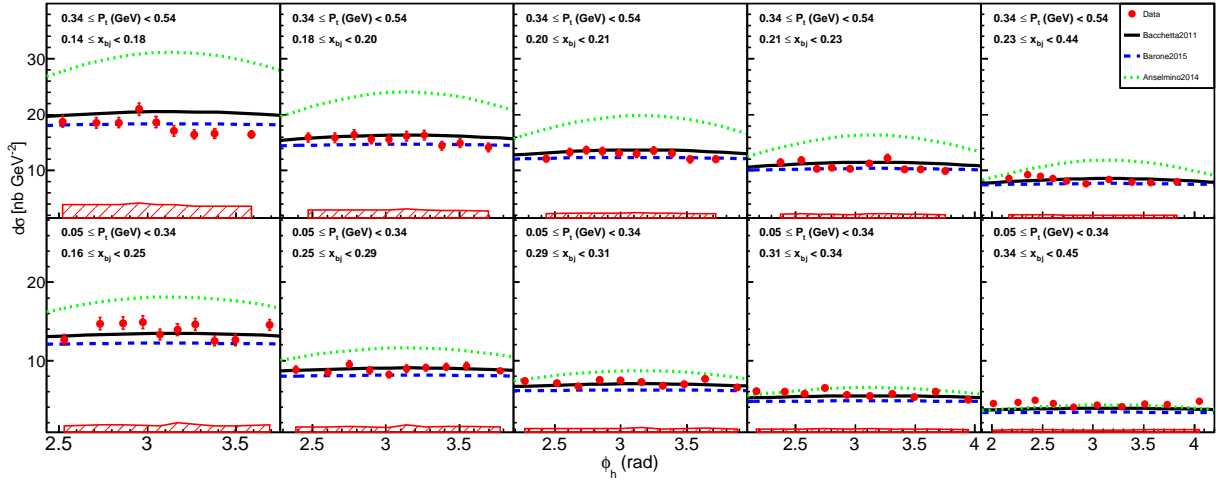}
\caption{(color online). 3D binning cross section for $\pi^+$  channel. The red circles are from the data, the black solid lines are from the model \cite{Bacchetta2011}, the blue dashed lines are from the model \cite{Barone2015} and the green dotted lines are from the model \cite{Anselmino2014}. The error bars represent the statistical uncertainties of the data. The error band on the bottom of each panel represents the experimental systematic uncertainty.}
\label{2510_3theo_pip_xs}
\end{center}
\end{figure*}

\begin{figure*}
\begin{center}
\includegraphics[width=0.9\textwidth]{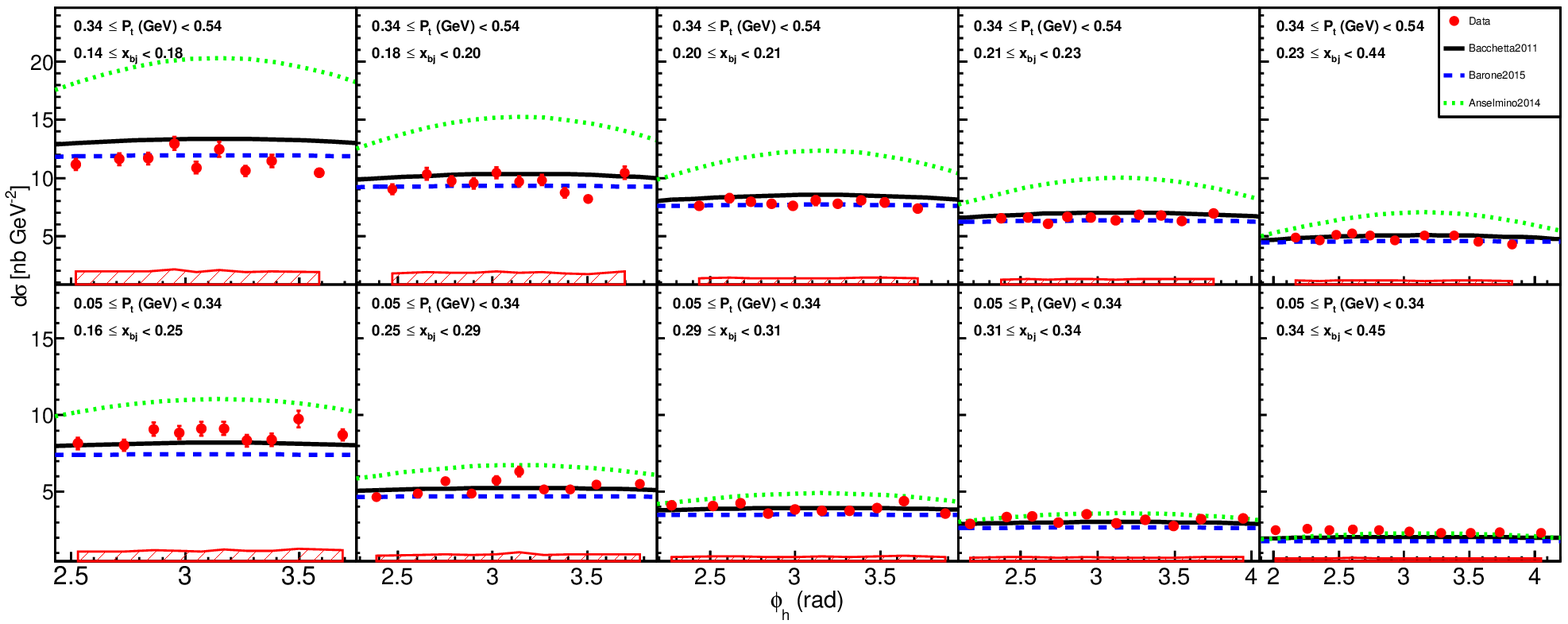}
\caption{(color online). 3D binning cross section for $\pi^-$  channel. The definitions of the markers, the lines and the bands are the same as the figure above for the $\pi^+$  channel.}
\label{2510_3theo_pim_xs}
\end{center}
\end{figure*}

\begin{figure*}
\begin{center}
\includegraphics[width=0.9\textwidth]{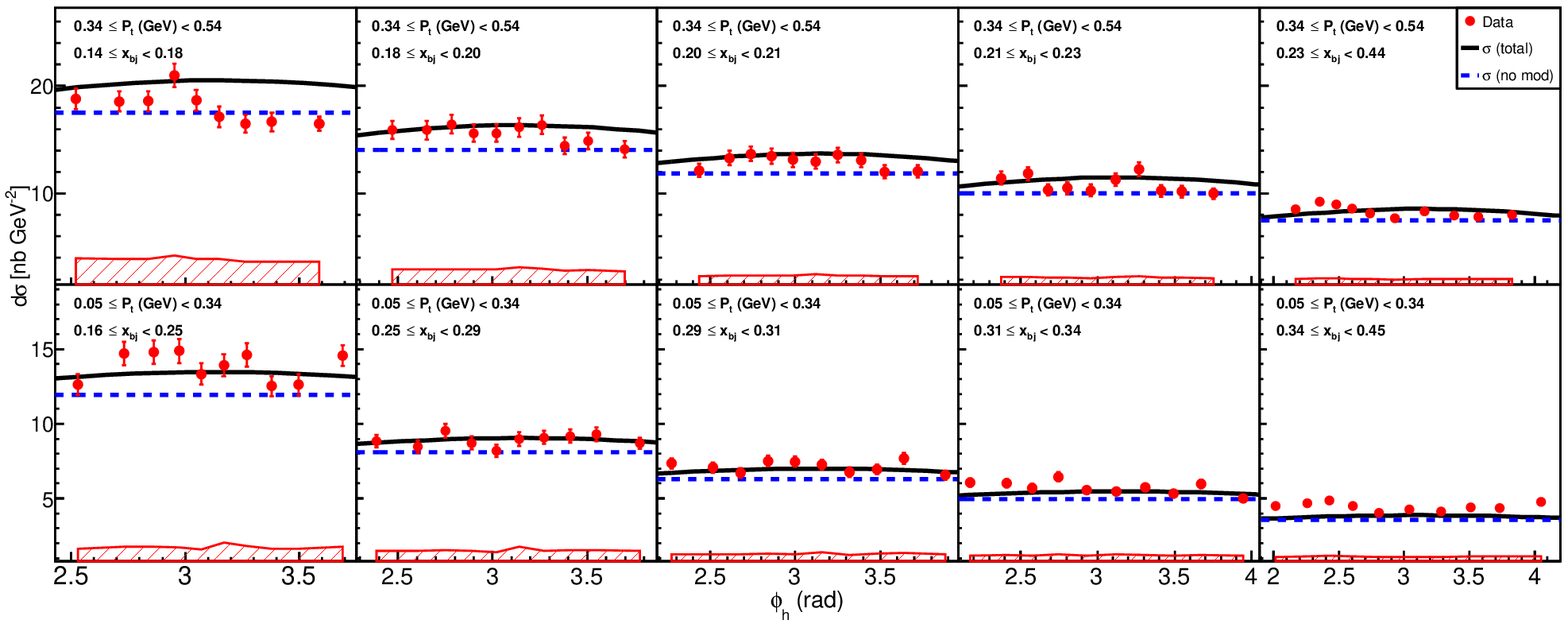}
\caption{(color online). 3D binning cross section for $\pi^+$  channel. The red circles are from the data, the black solid lines are from the model \cite{Bacchetta2011}, the blue dashed lines are from the model \cite{Bacchetta2011} with $F_{UU}^{\cos\phi_h}$ and $F_{UU}^{\cos2\phi_h}$ setting to zero. The error bars represent the statistical uncertainties of the data. The error band on the bottom of each panel represents the experimental systematic uncertainty.}
\label{2510_2011_mod_nomod_pip_xs}
\end{center}
\end{figure*}

\begin{figure*}
\begin{center}
\includegraphics[width=0.9\textwidth]{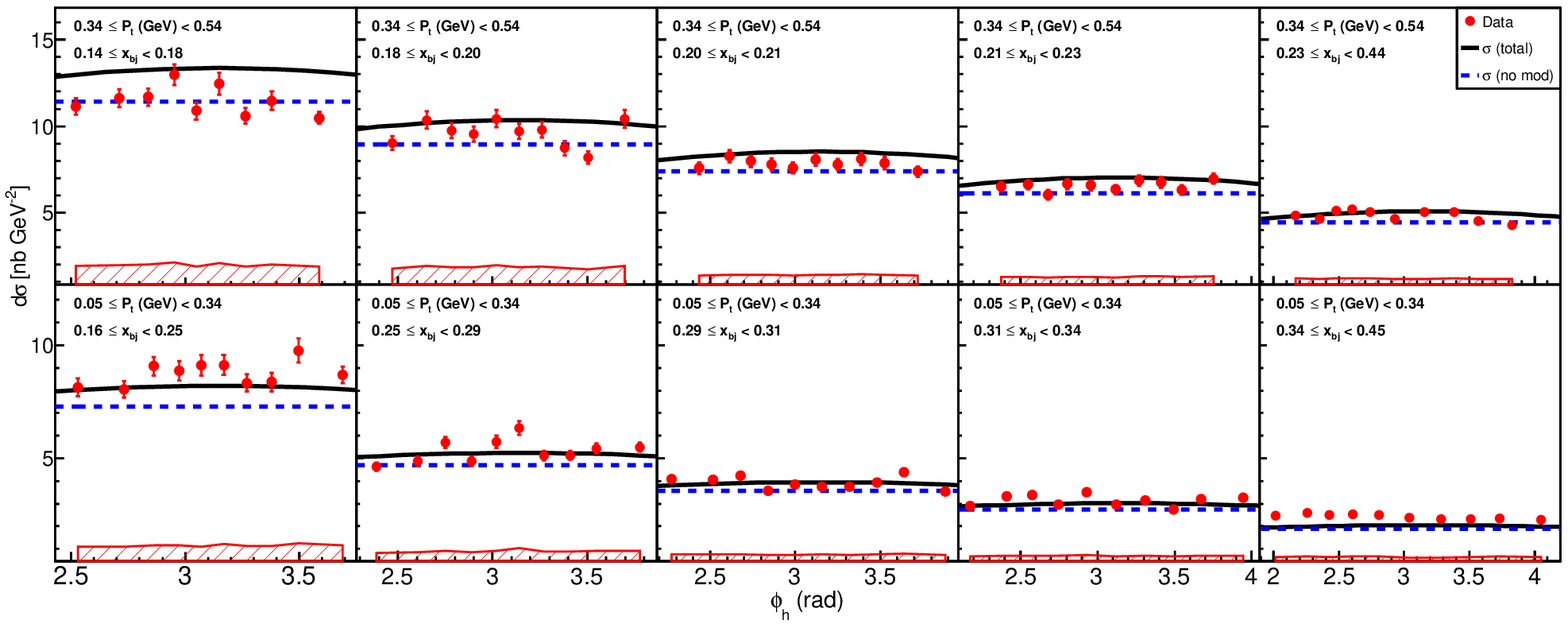}
\caption{(color online). 3D binning cross section for $\pi^-$  channel. The definitions of the markers, the lines and the bands are the same as the figure above for the $\pi^+$  channel.}
\label{2510_2011_mod_nomod_pim_xs}
\end{center}
\end{figure*}

The differential cross sections of SIDIS were extracted in 3D bins (2 $\times$ 5 $\times$ 10), to examine the $\phi_h$ dependence of the cross sections in 2 $\times$ 5 ranges of $P_t$ vs. $x_{bj}$. Bin-centering corrections were used to remove the difference of all the variables except $\phi_h$ from one bin to another in each of the $P_t$ vs. $x_{bj}$ ranges, therefore the 10 $\phi_h$ bins in a certain range of $P_t$ and $x_{bj}$ differ only in the values of $\phi_h$. The comparisons of the SIDIS differential cross sections from the data and the models from \cite{Barone2015, Anselmino2014, Bacchetta2011} are presented in Figs. \ref{2510_3theo_pip_xs} and \ref{2510_3theo_pim_xs}. Comparisons between the data and the model from \cite{Bacchetta2011} with and without modulation are in Figs. \ref{2510_2011_mod_nomod_pip_xs} and \ref{2510_2011_mod_nomod_pim_xs}. The comparisons show that the model from \cite{Bacchetta2011} compares the best with the data, while the model from \cite{Anselmino2014} deviates the most from the data in most of the kinematic ranges.

The cross sections and corresponding kinematic variables are recorded in Tables. \ref{xs_table_2510_p} and \ref{xs_table_2510_m} in Appendix \ref{xs-table}.

\subsection{The $P_t$ dependence of the cross sections}
\begin{figure*}
\begin{center}
\includegraphics[width=0.9\textwidth]{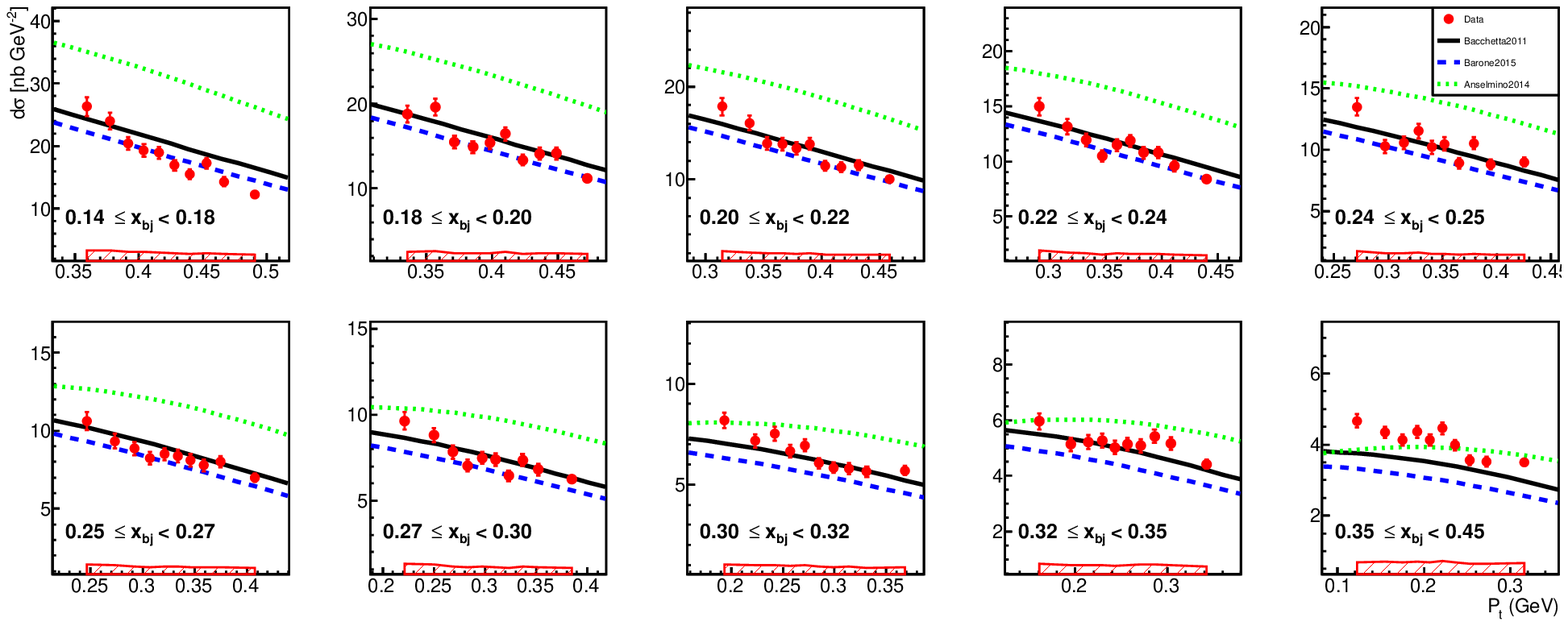}
\caption{(color online). 2D binning cross section for $\pi^+$  channel. The red circles are from the data, the black solid lines are from the model \cite{Bacchetta2011}, the blue dashed lines are from the model \cite{Barone2015} and the green dotted lines are from the model \cite{Anselmino2014}. The error bars represent the statistical uncertainties of the data. The error band on the bottom of each panel represents the experimental systematic uncertainty.}
\label{1010_3theo_pip_xs}
\end{center}
\end{figure*}

\begin{figure*}
\begin{center}
\includegraphics[width=0.9\textwidth]{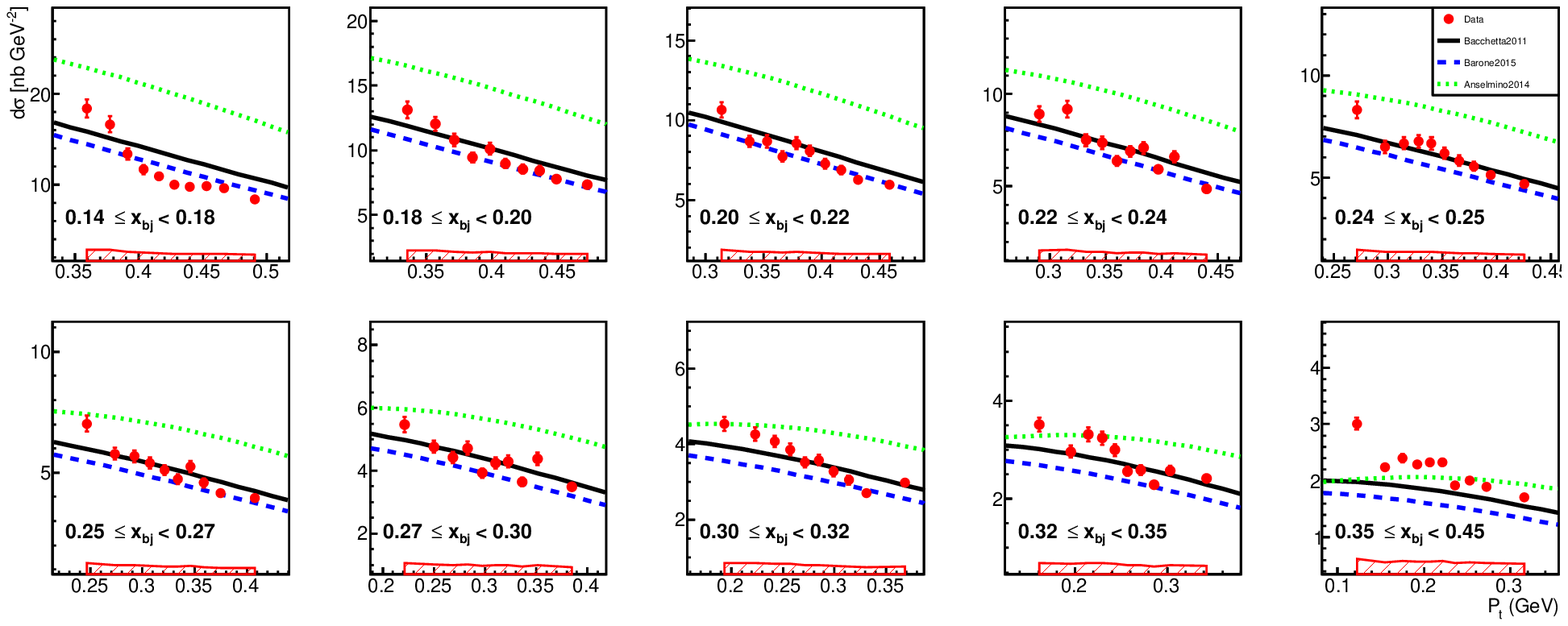}
\caption{(color online). 2D binning cross section for $\pi^-$  channel. The definitions of the markers, the lines and the bands are the same as the figure above for $\pi^+$  channel.}
\label{1010_3theo_pim_xs}
\end{center}
\end{figure*}

To present the $P_t$ dependence of the SIDIS cross sections, 2D bins (10 $\times$ 10) of $x_{bj}$ vs. $P_t$ are used. Bin-centering corrections were used to remove the difference of all the variables except $P_t$ from one bin to another in each range of $x_{bj}$, therefore the 10 $P_t$ bins in a certain range of $x_{bj}$ differ only in the values of $P_t$. The comparisons of the SIDIS differential cross sections from the data and the models from \cite{Barone2015, Anselmino2014, Bacchetta2011} are presented in Figs. \ref{1010_3theo_pip_xs} and \ref{1010_3theo_pim_xs}. The comparisons show that the model from \cite{Bacchetta2011} compares the best with the data, while the model from \cite{Anselmino2014} deviates the most from the data in most of the kinematic ranges. In the highest $x_{bj}$ ranges (corresponding to lowest $P_t$ ranges), the model from \cite{Anselmino2014} gives better comparison than the models from \cite{Barone2015,Bacchetta2011}, but still has sizable deviations from the data.

The cross sections and corresponding kinematic variables are recorded in Tables. \ref{xs_table_1010_p} and \ref{xs_table_1010_m} in Appendix \ref{xs-table}.

\subsection{The ratios of cross sections}

\begin{figure}
\begin{center}
\includegraphics[width=0.45\textwidth]{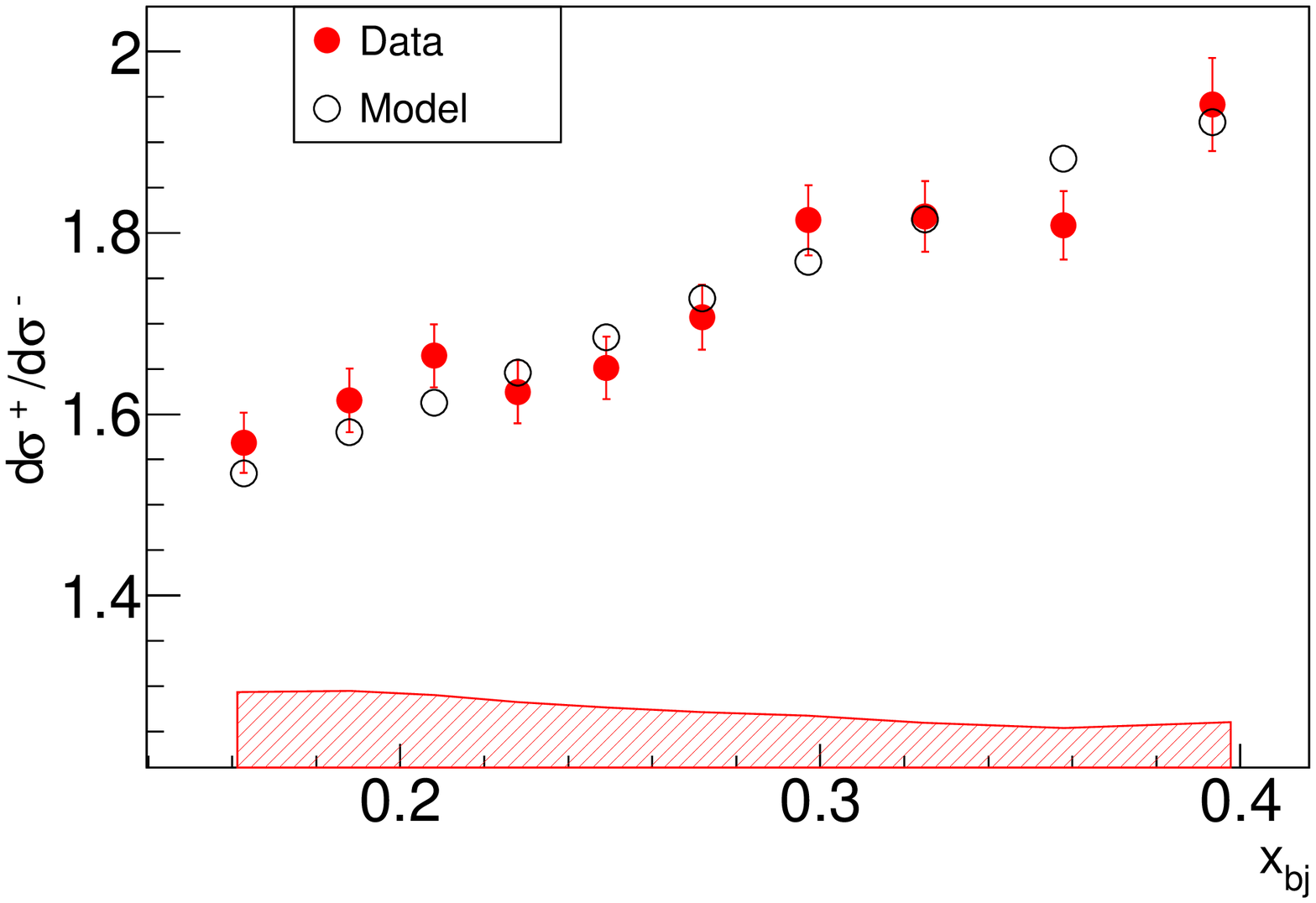}
\caption{(color online). SIDIS differential cross section ratio $\sigma^{\pi^+}$/$\sigma^{\pi^-}$ comparison in pseudo-1D $x_{bj}$ bins. The red solid circles are from the data and the black open circles are from the quark-parton model. The error bar of each point of data represents the statistical uncertainty. The error bars for the model parameterization uncertainty are smaller than the marker size. The error band on the bottom represents the systematic uncertainty of the data.}
\label{result_xs_ratio_x}
\end{center}
\end{figure}

The comparisons of the ratios (from the data and model) of SIDIS $\pi^+$ production cross sections over SIDIS $\pi^-$ production cross sections in pseudo-1D $x_{bj}$ bins are shown in Fig. \ref{result_xs_ratio_x}. The model parameters are the same as in part \ref{xs-1d} of this section. The systematic uncertainties from the acceptance and efficiency of electron detection in BigBite, are not included in the bottom systematic error band, as the electron part is the same in the SIDIS $\pi^{\pm}$ production.

In the plot, the error bars of the data points are for the statistical uncertainties of the data. The error bars of the model points are for the model uncertainties. In this study, the model uncertainties are defined by the quadrature combination of the differences of the ratios with and without the contribution from the Boer-Mulders terms, changing the width $\langle k_{\perp}^2 \rangle$ to $2\langle k_{\perp}^2 \rangle$ and changing $\langle p_{\perp}^2 \rangle$ to $2\langle p_{\perp}^2 \rangle$. The Boer-Mulders effects in the $\pi^{\pm}$ production channels have opposite signs, and the changes of the cross section ratios due to turning off the Boer-Mulders contributions are 1$\%$ to 4$\%$.  The flavor dependence of the widths has not been included in the model, thus the widths do not differ in channels of the $\pi^{\pm}$ production. Theoretically, if the $\pi^{\pm}$ production SIDIS cross sections have the same transverse momentum dependence, their ratios at the same kinematics will be independent of the widths. Due to very small differences between the central values of variables in the $\pi^{\pm}$ production channels, the effect of changing $\langle k_{\perp}^2 \rangle$ to $2\langle k_{\perp}^2 \rangle$ or $\langle p_{\perp}^2 \rangle$ to $2\langle p_{\perp}^2 \rangle$ was none-zero but less than 0.1$\%$.

Results from the data are consistent with the model without a flavor dependence of $\langle k_{\perp}^2 \rangle$ and $\langle p_{\perp}^2 \rangle$ as assumed in most of the global analysis for SIDIS \cite{Barone2015, Anselmino2014, Bacchetta2011}.

\subsection{Azimuthal modulation and stand-alone data fitting}
\begin{figure*}
\begin{center}
\includegraphics[width=0.9\textwidth]{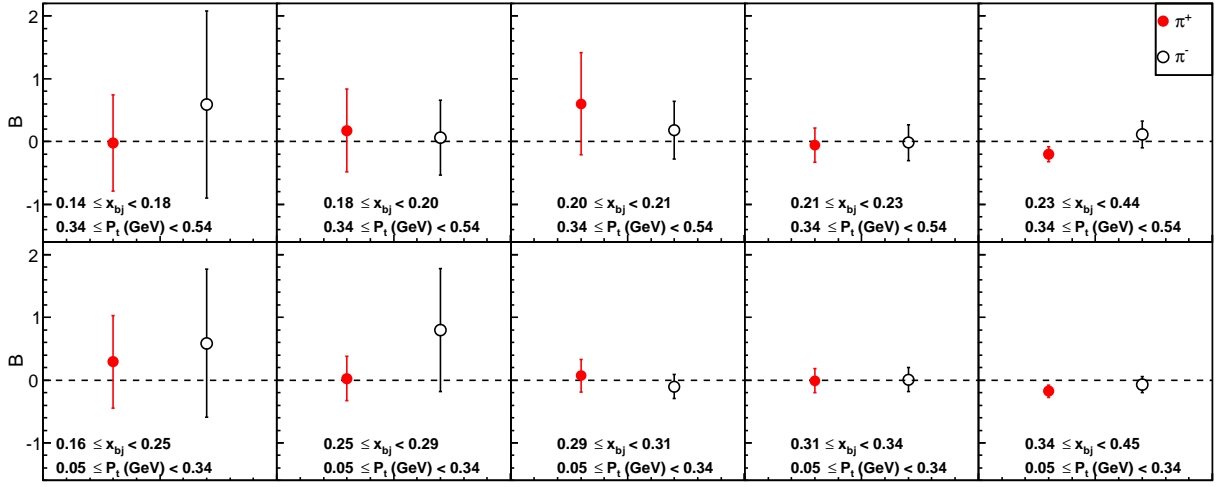}
\caption{(color online). Parameter $B$'s in 3D binning $A\cdot(1-B\cdot\cos\phi_h)$ fit.}
\label{2510_Bmod}
\end{center}
\end{figure*}

Fitting the $\phi_h$ distribution in each of the 2 $\times$ 5 ranges of $P_t$ vs. $x_{bj}$ in the 3D bins (2 $\times$ 5 $\times$ 10), with a simple function $A\cdot(1-B\cdot\cos\phi_h)$, provides a naive probe for the azimuthal modulation effect in the data. The parameter $B$ indicates the size of the modulation. The parameter $B$s in all ranges are presented in Fig. \ref{2510_Bmod}. Due to a limited $\phi_h$ range in the data and a large number of fitting parameters being used ($A$ and $B$ in one $P_t$ and $x_{bj}$ range differ from $A$ and $B$ in another range), the data do not provide good constraints on the $B$s.

Azimuthal modulation effects in the unpolarized SIDIS channel arise from the relative magnitudes of
$F_{UU}^{\cos\phi_h}$, $F_{UU}^{\cos2\phi_h}$ and $F_{UU}$. Using the functional forms in section \ref{sec-theory}, $F_{UU}$ and the Cahn parts of the structure functions $F_{UU}^{\cos\phi_h}$ and $F_{UU}^{\cos2\phi_h}$ after convolution can be expressed as
\begin{eqnarray} \label{after_conv}
F_{UU} &=& \sum_q f_q^c D_q^c \frac{e_q^2 x_{bj}}{\pi \langle P_t^2 \rangle} e^{-P_t^2/ \langle P_t^2 \rangle}, \\
F_{UU}^{\cos\phi_h} &=& -2 \sum_q f_q^c D_q^c \frac{P_t z_h e_q^2 x_{bj} \langle k_{\perp}^2 \rangle}{\pi Q \langle P_t^2 \rangle^2} e^{-P_t^2/ \langle P_t^2 \rangle}, \\
F_{UU}^{\cos2\phi_h} &=& 2 \sum_q f_q^c D_q^c \frac{P_t^2 z_h^2 e_q^2 x_{bj} \langle k_{\perp}^2 \rangle^2}{\pi Q^2 \langle P_t^2 \rangle^3} e^{-P_t^2/ \langle P_t^2 \rangle},
\end{eqnarray}
where $\langle P_t^2 \rangle = \langle p_{\perp}^2 \rangle + z_{h}^2 \langle k_{\perp}^2 \rangle$. The Boer-Mulders parts after convolution can be found in \cite{Barone2015}.

\begin{figure}
\begin{center}
\includegraphics[width=0.45\textwidth]{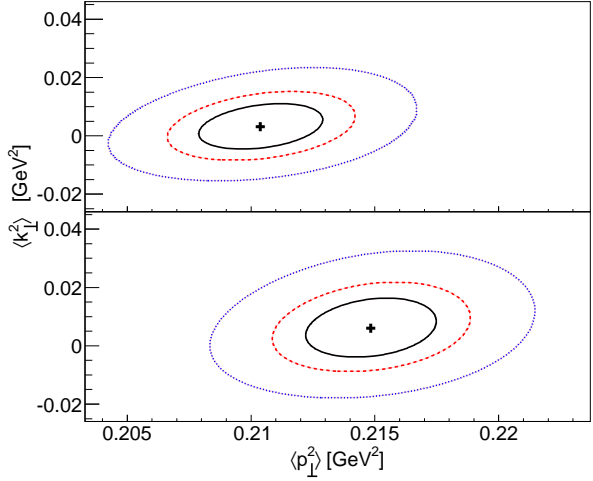}
\caption{(color online). Fitting contours with total unpolarized SIDIS cross section (refer to the text). The top panel is for the fitting results using the 2D bins (10 $\times$ 10) data, the bottom panel for the 3D bins (2 $\times$ 5 $\times$ 10). The central values of the fitting are the black crosses. The three contours from the smallest to the largest in each panel correspond to $\delta \chi^2=$ 1, 2.3 and 6.2, respectively.}
\label{siguu_dchi2}
\end{center}
\end{figure}

The parameters being fitted are the Gaussian widths $\langle k_{\perp}^2 \rangle$ and $\langle p_{\perp}^2 \rangle$, while the Boer-Mulders parts set to zero. The 2D bins (10 $\times$ 10) and 3D bins (2 $\times$ 5 $\times$ 10) data were fitted and the results are in Fig. \ref{siguu_dchi2}. Three contours corresponding to $\delta \chi^2=$ 1, 2.3 and 6.2 are drawn besides the central values from the fitting. $\delta \chi^2=$ 1 contour is conventionally the same as the 1-$\sigma$ contour. The contours of $\delta \chi^2=$ 2.3 and 6.2 show the constraints of two-parameter fitting at confidence levels of 68$\%$ and 95$\%$, respectively. The central values of the fitting in the 2D bins are $\langle k_{\perp}^2 \rangle=0.003\pm0.008$ GeV$^2$ and $\langle p_{\perp}^2 \rangle=0.2104\pm0.0025$ GeV$^2$. The central values of the fitting in the 3D bins are $\langle k_{\perp}^2 \rangle=0.006\pm0.010$ GeV$^2$ and $\langle p_{\perp}^2 \rangle=0.2148\pm0.0026$ GeV$^2$. The fitting results indicate consistent azimuthal modulation effects from the data in 3D bins with the $\phi_h$ information and 2D bins without the $\phi_h$ information.

\begin{figure}
\begin{center}
\includegraphics[width=0.45\textwidth]{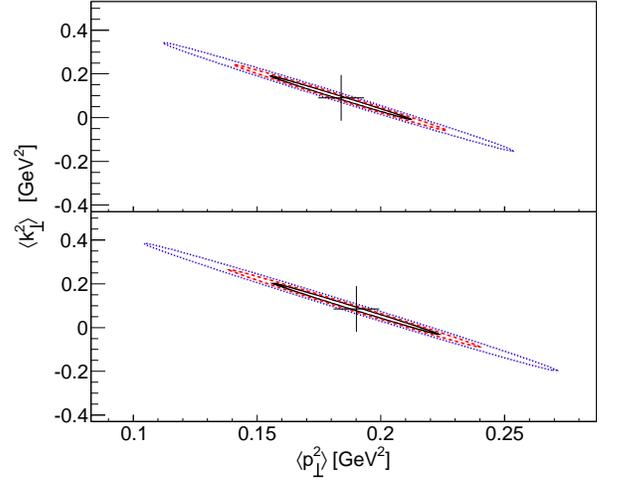}
\caption{(color online). Fitting contours with non-modulated unpolarized SIDIS cross section ($F_{UU}^{\cos\phi_h}$ and $F_{UU}^{\cos2\phi_h}$ set to zero: refer to the text). The top panel is for the fitting results using the 2D bins (10 $\times$ 10) data, the bottom panel for the 3D bins (2 $\times$ 5 $\times$ 10). The central values of the fitting are the black crosses. The three contours from the smallest to the largest in each panel correspond to $\delta \chi^2=$ 1, 2.3 and 6.2, respectively.}
\label{sig0_dchi2}
\end{center}
\end{figure}

Fitting the data with a simpler functional form, namely setting $F_{UU}^{\cos\phi_h}$ and $F_{UU}^{\cos2\phi_h}$ to zero, was also done. The results are presented in Fig. \ref{sig0_dchi2}. The central values of the fitting in the 2D bins are $\langle k_{\perp}^2 \rangle=0.090\pm0.097$ GeV$^2$ and $\langle p_{\perp}^2 \rangle=0.1840\pm0.0276$ GeV$^2$. The central values of the fitting in the 3D bins are $\langle k_{\perp}^2 \rangle=0.085\pm0.112$ GeV$^2$ and $\langle p_{\perp}^2 \rangle=0.1901\pm0.0330$ GeV$^2$.

The very different constraints of $\langle k_{\perp}^2 \rangle$ vs. $\langle p_{\perp}^2 \rangle$ using the functional form including all three structure functions (Fig. \ref{siguu_dchi2}) and the functional form including only structure function $F_{UU}$ (Fig. \ref{sig0_dchi2}), come from the specific model formulation, namely $F_{UU}^{\cos\phi_h}$ and $F_{UU}^{\cos2\phi_h}$ as in Eqs. (11) and (12). These specific functional forms, when applied to the data in this study, would result in the intrinsic transverse momentum width $\langle k_{\perp}^2 \rangle$ of the quarks in the nucleon being consistent with zero at small central values, which contradicts the results from the global analyses \cite{Barone2015, Anselmino2014, Bacchetta2011}. The effect of including the Boer-Mulders terms as parameterized in \cite{Barone2015} was tested to be negligible (less than 2$\%$ in the kinematic range of this study).

To examine the data's constraint on the intrinsic widths with relaxed model formulations, two adjusted functional forms were used to do the fitting in the 3D bins with the $\phi_h$ information. The first one includes the structure functions $F_{UU}$ and $F_{UU}^{\cos\phi_h}$, with an additional fitting parameter $A$ to tune the amplitude of modulation as $A\cdot F_{UU}^{\cos\phi_h}$. The results of the fitting are $\langle k_{\perp}^2 \rangle=0.078\pm0.1505$ GeV$^2$, $\langle p_{\perp}^2 \rangle=0.1925\pm0.0464$ GeV$^2$ and $A=0.0119\pm0.1971$. The intrinsic widths $\langle k_{\perp}^2 \rangle$ and $\langle p_{\perp}^2 \rangle$ are under loose constraint individually while the amplitude of $A\cdot F_{UU}^{\cos\phi_h}$ is suppressed by a small factor $A$.

The second one includes the structure functions $F_{UU}$, $F_{UU}^{\cos\phi_h}$ and $F_{UU}^{\cos2\phi_h}$, with an additional fitting parameter $A$ to tune the amplitude of modulation as $A\cdot( F_{UU}^{\cos\phi_h}+F_{UU}^{\cos2\phi_h})$. The results of the fitting are $\langle k_{\perp}^2 \rangle=0.080\pm0.1542$ GeV$^2$, $\langle p_{\perp}^2 \rangle=0.1918\pm0.0475$ GeV$^2$ and $A=0.0077\pm0.1820$. The intrinsic widths are under similar constraint as in the first case with a small factor $A$ suppressing the amplitude of $A\cdot( F_{UU}^{\cos\phi_h}+F_{UU}^{\cos2\phi_h})$.

\begin{figure*}
\begin{center}
\includegraphics[width=0.9\textwidth]{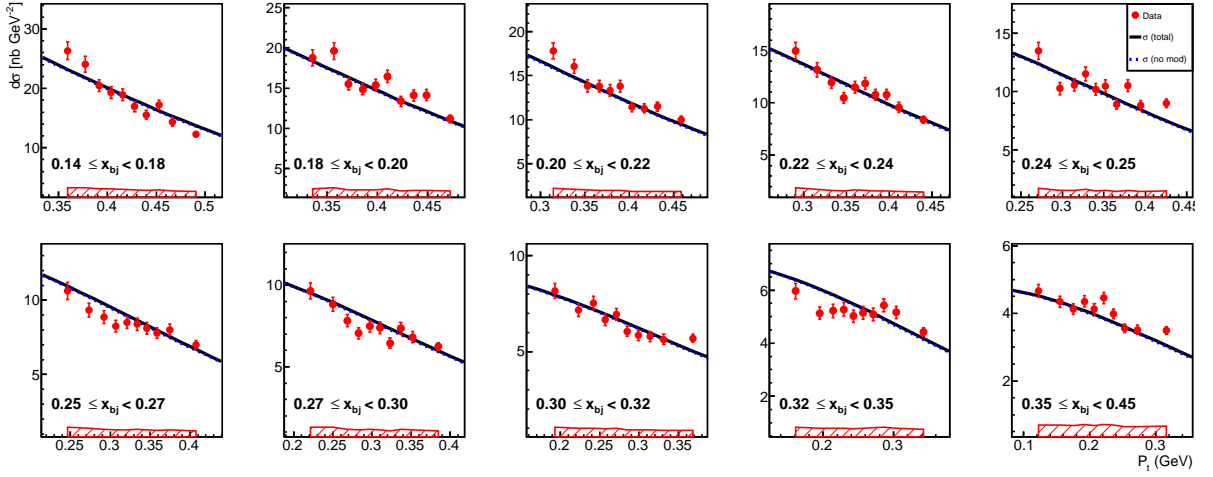}
\caption{(color online). 2D binning cross section for $\pi^+$  channel. The red circles are from the data. The black solid lines are from the model including the structure functions $F_{UU}$, $F_{UU}^{\cos\phi_h}$ and $F_{UU}^{\cos2\phi_h}$ with parameters $\langle k_{\perp}^2 \rangle$ and $\langle p_{\perp}^2 \rangle$ from stand-alone data fitting. The blue dashed lines are from the model including only the structure functions $F_{UU}$ with parameters $\langle k_{\perp}^2 \rangle$ and $\langle p_{\perp}^2 \rangle$ from fitting the data of this work only. The error bars represent the statistical uncertainties of the data. The error band on the bottom of each panel represents the experimental systematic uncertainty.}
\label{1010_bfs_pip_xs}
\end{center}
\end{figure*}

\begin{figure*}
\begin{center}
\includegraphics[width=0.9\textwidth]{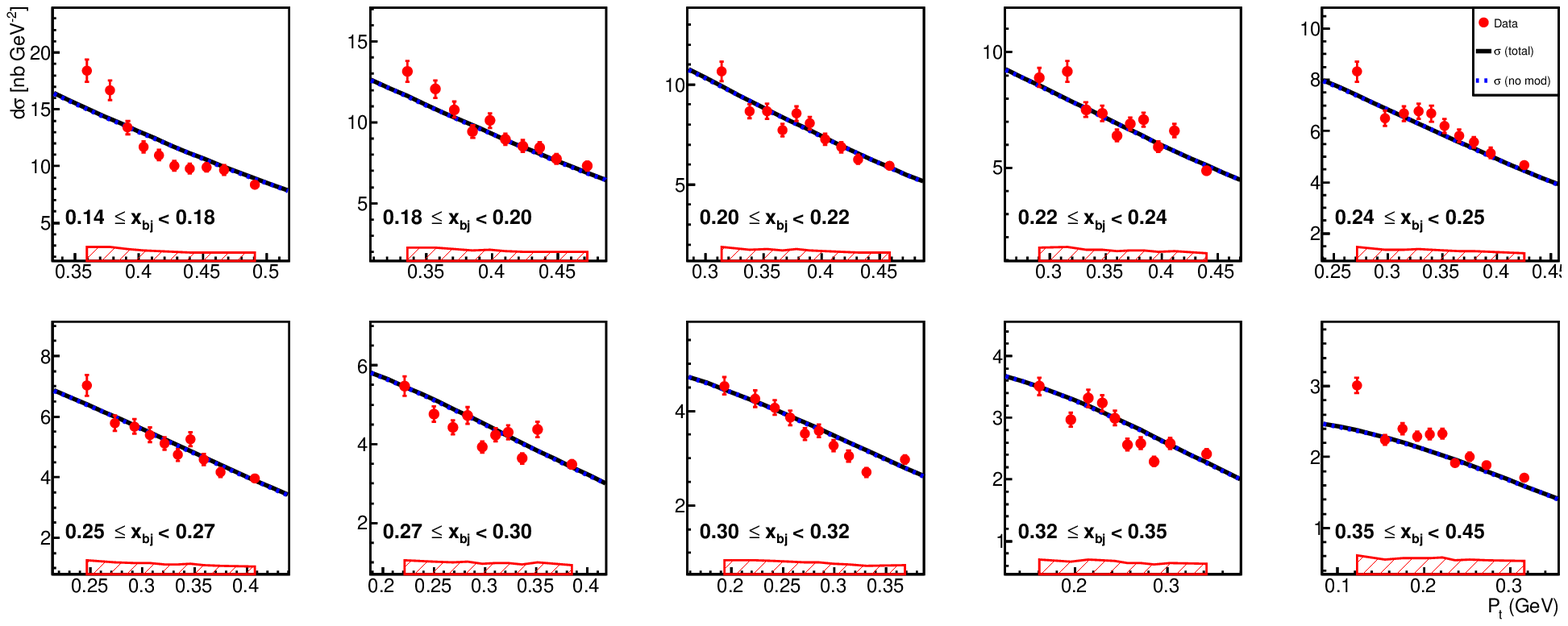}
\caption{(color online). 2D binning cross section for $\pi^-$  channel. The definitions of the markers, the lines and the bands are the same as the figure above for $\pi^+$  channel.}
\label{1010_bfs_pim_xs}
\end{center}
\end{figure*}

Without introducing specific forms of $F_{UU}^{\cos\phi_h}$ and $F_{UU}^{\cos2\phi_h}$, the parameters $\langle k_{\perp}^2 \rangle$ and $\langle p_{\perp}^2 \rangle$ in the SIDIS channels appear as the combined quantity $\langle P_t^2 \rangle$. Sensitivity to $\langle P_t^2 \rangle$ is explicitly provided by the $P_t$ behavior of the data. The comparison between the data and the models in the two functional forms (with and without $F_{UU}^{\cos\phi_h}$ and $F_{UU}^{\cos2\phi_h}$) using the parameters from fitting in 2D bins of the data are shown in Figs. \ref{1010_bfs_pip_xs} and \ref{1010_bfs_pim_xs}.

\section{Conclusion}

We report the first measurement of the unpolarized SIDIS differential cross section of $\pi^{\pm}$ production from a $^3$He target in a kinematic range $0.12 < x_{bj} < 0.45$, $1 < Q^2 < 4 \,$(GeV/c)$^2$, $0.45 < z_{h} < 0.65$, and $0.05 < P_t < 0.55 \,$GeV/c. Results in multi-dimensional bin sets show that the model from \cite{Bacchetta2011} compares the best with the data when approximating the $^3$He nucleus as two protons and one neutron (PWIA) in the SIDIS processes. The existing models \cite{Barone2015, Anselmino2014, Bacchetta2011} in the framework of naive $x$-$z$ factorization and fitting different types of global data (multiplicities and/or asymmetries), deviate from our results to certain extents in different kinematic ranges.

Azimuthal modulations in unpolarized SIDIS are observed to be consistent with zero within the experimental uncertainties in this study. Using the specific functional form as in the global analysis \cite{Barone2015}, the fitting results show that the width of quark intrinsic transverse momentum $\langle k_{\perp}^2 \rangle$ is much smaller than the results from the global analyses of other types of data \cite{Barone2015, Anselmino2014, Bacchetta2011}. With relaxed model formulation, $\langle k_{\perp}^2 \rangle$ and $\langle p_{\perp}^2 \rangle$ are under looser constraint individually, while the combined quantity $\langle P_t^2 \rangle$ is constrained by the $P_t$ behavior of the data. The widths $\langle k_{\perp}^2 \rangle$ and $\langle p_{\perp}^2 \rangle$ in the structure functions, related to the azimuthal modulations are determined consistently using the extracted cross sections with and without the information of $\phi_h$. 

It is foreseeable that the simple formulation (naive $x$-$z$ factorization) based on simplified TMD factorization and the Gaussian ansatz needs to be updated to describe multiple types of data (multiplicities, asymmetries and cross sections) in different kinematic ranges, at the same time.

The future 12 GeV SIDIS programs at JLab with SoLID combining high luminosities and a large acceptance including a full azimuthal coverage \cite{Jlab12GeV,SoLID_white_paper} will provide opportunities for precision measurements of the SIDIS differential cross sections as well as the azimuthal modulations in a broad kinematic range. These data will significantly benefit the developments of the TMD phenomenology and models.

\section{Acknowledgements}

We acknowledge the outstanding support of the JLab Hall
A staff and the Accelerator Division in accomplishing
this experiment. This work was supported in part by the
U. S. National Science Foundation, and by Department of Energy (DOE)
contract number DE-AC05-06OR23177, under which the Jefferson Science
Associates operates the Thomas Jefferson National Accelerator
Facility. This work was also supported in part by the U.S. Department of Energy under Contract DE-FG02-03ER41231 and the National Natural Science Foundation of China No. 11120101004.


\appendix
\section{\label{acc-corr} Acceptance correction}
The numbers of events in a specific bin of data, weighted and non-weighted simulations are expressed as
\begin{eqnarray} \label{acc-corr-eqs}
N_{data} &=& \overline{\frac{d\sigma}{dPHS}}_{d} \cdot \Delta PHS_d \cdot f_{acc,d} \cdot L_d ,\\
N_{sim} &=& \overline{\frac{d\sigma}{dPHS}}_{s} \cdot \Delta PHS_s \cdot f_{acc,s} \cdot L_s ,\\
N_{phs} &=& 1 \cdot \Delta PHS_s \cdot f_{acc,s} \cdot L_s,
\end{eqnarray}
where $N_{data}$ is the number of events from the data, $N_{sim}$ from the weighted simulation and $N_{phs}$ from the non-weighted simulation. $\overline{\frac{d\sigma}{dPHS}}_{d}$ is the (averaged) differential cross section from the data in a specific bin, and $\overline{\frac{d\sigma}{dPHS}}_{s}$ from the theory in the weighted simulation. $\Delta PHS_d$ is the total phase space in the data for a specific bin, and $\Delta PHS_s$ in the simulations. $f_{acc,d}$ is the acceptance factor in the data, and $f_{acc,s}$ in the simulations. $L_d$ is the luminosity in the data, and $L_s$ in the simulations. The data luminosity $L_d$ is the product of the total number of beam-electrons (from the recorded beam charge), the live-time of the data-acquisition system and the target thickness (from the target density and the target length).

The description of experimental acceptance can be tested by using known channels, where $\overline{\frac{d\sigma}{dPHS}}_{d}=\overline{\frac{d\sigma}{dPHS}}_{s}$, with proper scaling of the luminosities. The tested framework of simulation provides $\Delta PHS_s=\Delta PHS_d$ and $f_{acc,s}=f_{acc,d}$. The data have to be corrected by all the other data corrections for meaningful comparisons of the numbers of events $N_{data}$ and $N_{sim}$ in each bin.

The (averaged) differential cross section from the theory can be retrieved by $N_{sim}/N_{phs}$, and the cross section extracted from the data is expressed as $(N_{data}/N_{phs})\times (L_s/L_d)$.

\section{\label{xs-table} SIDIS cross section data table}
The SIDIS differential cross sections in 2D and 3D bins and the corresponding kinematic variables are presented in the tables below.

\begin{table*}[ht]
\caption{Unpolarized SIDIS cross section in 3D bins for the $\pi^+$ channel. $Q^2$ is in unit of GeV$^2$, $P_t$ in unit of GeV, $\phi_h$ in unit of rad. $d\sigma$ stands for $d\sigma/(dx_{bj}dydz_{h} d\phi_S dP_t^2d\phi_h)$ in unit of $\rm{nb}\cdot \rm{GeV}^{-2}\cdot rad^{-2}$. Stat. unc. and Sys. unc. stand for statistical and systematic uncertainties, respectively, in unit of $\rm{nb}\cdot \rm{GeV}^{-2}\cdot rad^{-2}$.}
\begin{tabular}{|c c c c c c c c c|c c c c c c c c c|}\hline
$x_{bj}$ & $Q^2$ & y & $z_{h}$ & $P_t$ & $\phi_h$ &  $d\sigma$ & Stat. unc. & Sys. unc. & $x_{bj}$ & $Q^2$ & y & $z_{h}$ & $P_t$ & $\phi_h$ &  $d\sigma$ & Stat. unc. & Sys. unc. \\ \hline
0.226 & 1.95 & 0.783 & 0.504 & 0.312 & 2.53 & 12.62 & 0.68 & 0.83 & 0.226 & 1.95 & 0.783 & 0.504 & 0.312 & 2.73 & 14.72 & 0.78 & 0.98 \\
0.226 & 1.95 & 0.783 & 0.504 & 0.312 & 2.86 & 14.79 & 0.78 & 0.96 & 0.226 & 1.95 & 0.783 & 0.504 & 0.312 & 2.97 & 14.87 & 0.82 & 0.92 \\
0.226 & 1.95 & 0.783 & 0.504 & 0.312 & 3.07 & 13.33 & 0.71 & 0.79 & 0.226 & 1.95 & 0.783 & 0.504 & 0.312 & 3.17 & 13.93 & 0.73 & 1.2 \\
0.226 & 1.95 & 0.783 & 0.504 & 0.312 & 3.27 & 14.61 & 0.77 & 1 & 0.226 & 1.95 & 0.783 & 0.504 & 0.312 & 3.38 & 12.53 & 0.67 & 0.84 \\
0.226 & 1.95 & 0.783 & 0.504 & 0.312 & 3.5 & 12.61 & 0.71 & 0.85 & 0.226 & 1.95 & 0.783 & 0.504 & 0.312 & 3.69 & 14.56 & 0.7 & 0.98 \\ \hline
0.27 & 2.26 & 0.758 & 0.522 & 0.287 & 2.39 & 8.842 & 0.43 & 0.67 & 0.27 & 2.26 & 0.758 & 0.522 & 0.287 & 2.61 & 8.457 & 0.42 & 0.68 \\
0.27 & 2.26 & 0.758 & 0.522 & 0.287 & 2.75 & 9.518 & 0.47 & 0.76 & 0.27 & 2.26 & 0.758 & 0.522 & 0.287 & 2.89 & 8.713 & 0.43 & 0.69 \\
0.27 & 2.26 & 0.758 & 0.522 & 0.287 & 3.02 & 8.187 & 0.42 & 0.58 & 0.27 & 2.26 & 0.758 & 0.522 & 0.287 & 3.14 & 8.969 & 0.47 & 0.95 \\
0.27 & 2.26 & 0.758 & 0.522 & 0.287 & 3.27 & 9.084 & 0.43 & 0.7 & 0.27 & 2.26 & 0.758 & 0.522 & 0.287 & 3.41 & 9.17 & 0.44 & 0.74 \\
0.27 & 2.26 & 0.758 & 0.522 & 0.287 & 3.55 & 9.306 & 0.47 & 0.72 & 0.27 & 2.26 & 0.758 & 0.522 & 0.287 & 3.78 & 8.685 & 0.37 & 0.68 \\ \hline
0.299 & 2.45 & 0.741 & 0.534 & 0.267 & 2.27 & 7.354 & 0.35 & 0.46 & 0.299 & 2.45 & 0.741 & 0.534 & 0.267 & 2.52 & 7.077 & 0.34 & 0.45 \\
0.299 & 2.45 & 0.741 & 0.534 & 0.267 & 2.68 & 6.695 & 0.32 & 0.44 & 0.299 & 2.45 & 0.741 & 0.534 & 0.267 & 2.84 & 7.507 & 0.35 & 0.5 \\
0.299 & 2.45 & 0.741 & 0.534 & 0.267 & 3 & 7.471 & 0.35 & 0.46 & 0.299 & 2.45 & 0.741 & 0.534 & 0.267 & 3.16 & 7.268 & 0.34 & 0.62 \\
0.299 & 2.45 & 0.741 & 0.534 & 0.267 & 3.32 & 6.759 & 0.33 & 0.43 & 0.299 & 2.45 & 0.741 & 0.534 & 0.267 & 3.48 & 6.93 & 0.33 & 0.5 \\
0.299 & 2.45 & 0.741 & 0.534 & 0.267 & 3.64 & 7.676 & 0.38 & 0.56 & 0.299 & 2.45 & 0.741 & 0.534 & 0.267 & 3.88 & 6.556 & 0.29 & 0.44 \\ \hline
0.327 & 2.63 & 0.726 & 0.546 & 0.243 & 2.17 & 6.066 & 0.29 & 0.38 & 0.327 & 2.63 & 0.726 & 0.546 & 0.243 & 2.41 & 6.014 & 0.29 & 0.42 \\
0.327 & 2.63 & 0.726 & 0.546 & 0.243 & 2.58 & 5.696 & 0.28 & 0.39 & 0.327 & 2.63 & 0.726 & 0.546 & 0.243 & 2.75 & 6.46 & 0.3 & 0.45 \\
0.327 & 2.63 & 0.726 & 0.546 & 0.243 & 2.93 & 5.579 & 0.28 & 0.38 & 0.327 & 2.63 & 0.726 & 0.546 & 0.243 & 3.12 & 5.458 & 0.26 & 0.45 \\
0.327 & 2.63 & 0.726 & 0.546 & 0.243 & 3.31 & 5.735 & 0.28 & 0.39 & 0.327 & 2.63 & 0.726 & 0.546 & 0.243 & 3.49 & 5.343 & 0.26 & 0.38 \\
0.327 & 2.63 & 0.726 & 0.546 & 0.243 & 3.67 & 5.981 & 0.29 & 0.43 & 0.327 & 2.63 & 0.726 & 0.546 & 0.243 & 3.95 & 4.986 & 0.21 & 0.35 \\ \hline
0.371 & 2.89 & 0.706 & 0.561 & 0.198 & 2.02 & 4.479 & 0.16 & 0.3 & 0.371 & 2.89 & 0.706 & 0.561 & 0.198 & 2.25 & 4.664 & 0.17 & 0.31 \\
0.371 & 2.89 & 0.706 & 0.561 & 0.198 & 2.43 & 4.881 & 0.17 & 0.35 & 0.371 & 2.89 & 0.706 & 0.561 & 0.198 & 2.61 & 4.518 & 0.16 & 0.33 \\
0.371 & 2.89 & 0.706 & 0.561 & 0.198 & 2.81 & 4.022 & 0.15 & 0.28 & 0.371 & 2.89 & 0.706 & 0.561 & 0.198 & 3.04 & 4.278 & 0.15 & 0.29 \\
0.371 & 2.89 & 0.706 & 0.561 & 0.198 & 3.28 & 4.116 & 0.15 & 0.27 & 0.371 & 2.89 & 0.706 & 0.561 & 0.198 & 3.51 & 4.402 & 0.15 & 0.33 \\
0.371 & 2.89 & 0.706 & 0.561 & 0.198 & 3.73 & 4.355 & 0.15 & 0.33 & 0.371 & 2.89 & 0.706 & 0.561 & 0.198 & 4.05 & 4.764 & 0.13 & 0.34 \\ \hline
0.163 & 1.49 & 0.826 & 0.48 & 0.427 & 2.52 & 18.78 & 0.96 & 2.4 & 0.163 & 1.49 & 0.826 & 0.48 & 0.427 & 2.71 & 18.58 & 0.95 & 2.4 \\
0.163 & 1.49 & 0.826 & 0.48 & 0.427 & 2.84 & 18.59 & 0.95 & 2.4 & 0.163 & 1.49 & 0.826 & 0.48 & 0.427 & 2.95 & 21.01 & 1.1 & 2.7 \\
0.163 & 1.49 & 0.826 & 0.48 & 0.427 & 3.05 & 18.68 & 0.99 & 2.4 & 0.163 & 1.49 & 0.826 & 0.48 & 0.427 & 3.15 & 17.14 & 0.94 & 2.3 \\
0.163 & 1.49 & 0.826 & 0.48 & 0.427 & 3.26 & 16.49 & 0.83 & 2.1 & 0.163 & 1.49 & 0.826 & 0.48 & 0.427 & 3.38 & 16.66 & 0.87 & 2.1 \\
0.163 & 1.49 & 0.826 & 0.48 & 0.427 & 3.59 & 16.49 & 0.63 & 2.1 & 0.163 & 1.49 & 0.826 & 0.48 & 0.427 & 3.59 & 16.49 & 0.63 & 2.1 \\ \hline
0.187 & 1.69 & 0.816 & 0.486 & 0.404 & 2.47 & 15.93 & 0.83 & 1.4 & 0.187 & 1.69 & 0.816 & 0.486 & 0.404 & 2.65 & 15.89 & 0.85 & 1.4 \\
0.187 & 1.69 & 0.816 & 0.486 & 0.404 & 2.78 & 16.45 & 0.87 & 1.4 & 0.187 & 1.69 & 0.816 & 0.486 & 0.404 & 2.9 & 15.63 & 0.83 & 1.4 \\
0.187 & 1.69 & 0.816 & 0.486 & 0.404 & 3.02 & 15.61 & 0.81 & 1.4 & 0.187 & 1.69 & 0.816 & 0.486 & 0.404 & 3.14 & 16.14 & 0.85 & 1.6 \\
0.187 & 1.69 & 0.816 & 0.486 & 0.404 & 3.26 & 16.38 & 0.86 & 1.4 & 0.187 & 1.69 & 0.816 & 0.486 & 0.404 & 3.38 & 14.45 & 0.78 & 1.3 \\
0.187 & 1.69 & 0.816 & 0.486 & 0.404 & 3.51 & 14.89 & 0.77 & 1.3 & 0.187 & 1.69 & 0.816 & 0.486 & 0.404 & 3.7 & 14.13 & 0.76 & 1.2 \\ \hline
0.205 & 1.83 & 0.808 & 0.49 & 0.395 & 2.43 & 12.14 & 0.62 & 0.76 & 0.205 & 1.83 & 0.808 & 0.49 & 0.395 & 2.61 & 13.3 & 0.67 & 0.81 \\
0.205 & 1.83 & 0.808 & 0.49 & 0.395 & 2.74 & 13.7 & 0.68 & 0.84 & 0.205 & 1.83 & 0.808 & 0.49 & 0.395 & 2.86 & 13.47 & 0.7 & 0.85 \\
0.205 & 1.83 & 0.808 & 0.49 & 0.395 & 2.99 & 13.12 & 0.66 & 0.8 & 0.205 & 1.83 & 0.808 & 0.49 & 0.395 & 3.12 & 12.99 & 0.69 & 0.97 \\
0.205 & 1.83 & 0.808 & 0.49 & 0.395 & 3.25 & 13.6 & 0.68 & 0.82 & 0.205 & 1.83 & 0.808 & 0.49 & 0.395 & 3.39 & 13.09 & 0.65 & 0.84 \\
0.205 & 1.83 & 0.808 & 0.49 & 0.395 & 3.52 & 12.01 & 0.62 & 0.76 & 0.205 & 1.83 & 0.808 & 0.49 & 0.395 & 3.72 & 12.06 & 0.58 & 0.74 \\ \hline
0.223 & 1.97 & 0.799 & 0.496 & 0.385 & 2.38 & 11.45 & 0.58 & 0.7 & 0.223 & 1.97 & 0.799 & 0.496 & 0.385 & 2.55 & 11.84 & 0.6 & 0.71 \\
0.223 & 1.97 & 0.799 & 0.496 & 0.385 & 2.68 & 10.32 & 0.54 & 0.61 & 0.223 & 1.97 & 0.799 & 0.496 & 0.385 & 2.81 & 10.5 & 0.53 & 0.62 \\
0.223 & 1.97 & 0.799 & 0.496 & 0.385 & 2.96 & 10.29 & 0.52 & 0.58 & 0.223 & 1.97 & 0.799 & 0.496 & 0.385 & 3.12 & 11.29 & 0.57 & 0.73 \\
0.223 & 1.97 & 0.799 & 0.496 & 0.385 & 3.27 & 12.24 & 0.65 & 0.73 & 0.223 & 1.97 & 0.799 & 0.496 & 0.385 & 3.41 & 10.24 & 0.53 & 0.64 \\
0.223 & 1.97 & 0.799 & 0.496 & 0.385 & 3.55 & 10.18 & 0.54 & 0.61 & 0.223 & 1.97 & 0.799 & 0.496 & 0.385 & 3.75 & 9.974 & 0.47 & 0.59 \\ \hline
0.256 & 2.21 & 0.782 & 0.507 & 0.37 & 2.16 & 8.558 & 0.31 & 0.53 & 0.256 & 2.21 & 0.782 & 0.507 & 0.37 & 2.35 & 9.222 & 0.33 & 0.55 \\
0.256 & 2.21 & 0.782 & 0.507 & 0.37 & 2.48 & 8.955 & 0.34 & 0.54 & 0.256 & 2.21 & 0.782 & 0.507 & 0.37 & 2.6 & 8.573 & 0.33 & 0.52 \\
0.256 & 2.21 & 0.782 & 0.507 & 0.37 & 2.74 & 8.131 & 0.31 & 0.48 & 0.256 & 2.21 & 0.782 & 0.507 & 0.37 & 2.93 & 7.682 & 0.29 & 0.46 \\
0.256 & 2.21 & 0.782 & 0.507 & 0.37 & 3.16 & 8.364 & 0.32 & 0.48 & 0.256 & 2.21 & 0.782 & 0.507 & 0.37 & 3.38 & 7.965 & 0.31 & 0.49 \\
0.256 & 2.21 & 0.782 & 0.507 & 0.37 & 3.57 & 7.827 & 0.29 & 0.48 & 0.256 & 2.21 & 0.782 & 0.507 & 0.37 & 3.83 & 7.988 & 0.23 & 0.5 \\ \hline
\end{tabular}
\label{xs_table_2510_p}
\end{table*}

\begin{table*}[ht]
\caption{Unpolarized SIDIS cross section in 3D bins for the $\pi^-$ channel. $Q^2$ is in unit of GeV$^2$, $P_t$ in unit of GeV, $\phi_h$ in unit of rad. $d\sigma$ stands for $d\sigma/(dx_{bj}dydz_{h} d\phi_S dP_t^2d\phi_h)$ in unit of $\rm{nb}\cdot \rm{GeV}^{-2}\cdot rad^{-2}$. Stat. unc. and Sys. unc. stand for statistical and systematic uncertainties, respectively, in unit of $\rm{nb}\cdot \rm{GeV}^{-2}\cdot rad^{-2}$.}
\begin{tabular}{|c c c c c c c c c|c c c c c c c c c|}\hline
$x_{bj}$ & $Q^2$ & y & $z_{h}$ & $P_t$ & $\phi_h$ &  $d\sigma$ & Stat. unc. & Sys. unc. & $x_{bj}$ & $Q^2$ & y & $z_{h}$ & $P_t$ & $\phi_h$ &  $d\sigma$ & Stat. unc. & Sys. unc. \\ \hline
0.225 & 1.95 & 0.784 & 0.504 & 0.312 & 2.53 & 8.15 & 0.4 & 0.63 & 0.225 & 1.95 & 0.784 & 0.504 & 0.312 & 2.73 & 8.051 & 0.36 & 0.63 \\
0.225 & 1.95 & 0.784 & 0.504 & 0.312 & 2.86 & 9.074 & 0.42 & 0.71 & 0.225 & 1.95 & 0.784 & 0.504 & 0.312 & 2.97 & 8.864 & 0.43 & 0.69 \\
0.225 & 1.95 & 0.784 & 0.504 & 0.312 & 3.07 & 9.121 & 0.45 & 0.64 & 0.225 & 1.95 & 0.784 & 0.504 & 0.312 & 3.17 & 9.126 & 0.44 & 0.76 \\
0.225 & 1.95 & 0.784 & 0.504 & 0.312 & 3.27 & 8.339 & 0.38 & 0.67 & 0.225 & 1.95 & 0.784 & 0.504 & 0.312 & 3.38 & 8.385 & 0.41 & 0.68 \\
0.225 & 1.95 & 0.784 & 0.504 & 0.312 & 3.5 & 9.762 & 0.54 & 0.8 & 0.225 & 1.95 & 0.784 & 0.504 & 0.312 & 3.69 & 8.691 & 0.37 & 0.71 \\ \hline
0.27 & 2.26 & 0.758 & 0.522 & 0.287 & 2.39 & 4.631 & 0.19 & 0.37 & 0.27 & 2.26 & 0.758 & 0.522 & 0.287 & 2.61 & 4.884 & 0.21 & 0.39 \\
0.27 & 2.26 & 0.758 & 0.522 & 0.287 & 2.75 & 5.697 & 0.25 & 0.47 & 0.27 & 2.26 & 0.758 & 0.522 & 0.287 & 2.89 & 4.874 & 0.2 & 0.4 \\
0.27 & 2.26 & 0.758 & 0.522 & 0.287 & 3.02 & 5.723 & 0.28 & 0.45 & 0.27 & 2.26 & 0.758 & 0.522 & 0.287 & 3.14 & 6.335 & 0.31 & 0.59 \\
0.27 & 2.26 & 0.758 & 0.522 & 0.287 & 3.27 & 5.131 & 0.21 & 0.42 & 0.27 & 2.26 & 0.758 & 0.522 & 0.287 & 3.41 & 5.135 & 0.21 & 0.44 \\
0.27 & 2.26 & 0.758 & 0.522 & 0.287 & 3.55 & 5.442 & 0.24 & 0.46 & 0.27 & 2.26 & 0.758 & 0.522 & 0.287 & 3.78 & 5.5 & 0.21 & 0.46 \\ \hline
0.299 & 2.45 & 0.741 & 0.534 & 0.268 & 2.27 & 4.09 & 0.17 & 0.29 & 0.299 & 2.45 & 0.741 & 0.534 & 0.268 & 2.52 & 4.044 & 0.17 & 0.29 \\
0.299 & 2.45 & 0.741 & 0.534 & 0.268 & 2.68 & 4.24 & 0.18 & 0.31 & 0.299 & 2.45 & 0.741 & 0.534 & 0.268 & 2.84 & 3.557 & 0.13 & 0.27 \\
0.299 & 2.45 & 0.741 & 0.534 & 0.268 & 3 & 3.836 & 0.15 & 0.27 & 0.299 & 2.45 & 0.741 & 0.534 & 0.268 & 3.16 & 3.763 & 0.15 & 0.29 \\
0.299 & 2.45 & 0.741 & 0.534 & 0.268 & 3.32 & 3.754 & 0.15 & 0.28 & 0.299 & 2.45 & 0.741 & 0.534 & 0.268 & 3.48 & 3.92 & 0.16 & 0.31 \\
0.299 & 2.45 & 0.741 & 0.534 & 0.268 & 3.64 & 4.379 & 0.18 & 0.35 & 0.299 & 2.45 & 0.741 & 0.534 & 0.268 & 3.88 & 3.552 & 0.13 & 0.27 \\ \hline
0.327 & 2.63 & 0.727 & 0.545 & 0.243 & 2.18 & 2.891 & 0.11 & 0.21 & 0.327 & 2.63 & 0.727 & 0.545 & 0.243 & 2.41 & 3.333 & 0.14 & 0.25 \\
0.327 & 2.63 & 0.727 & 0.545 & 0.243 & 2.58 & 3.389 & 0.14 & 0.26 & 0.327 & 2.63 & 0.727 & 0.545 & 0.243 & 2.75 & 2.97 & 0.11 & 0.23 \\
0.327 & 2.63 & 0.727 & 0.545 & 0.243 & 2.93 & 3.52 & 0.15 & 0.27 & 0.327 & 2.63 & 0.727 & 0.545 & 0.243 & 3.12 & 2.946 & 0.12 & 0.23 \\
0.327 & 2.63 & 0.727 & 0.545 & 0.243 & 3.31 & 3.143 & 0.13 & 0.24 & 0.327 & 2.63 & 0.727 & 0.545 & 0.243 & 3.49 & 2.744 & 0.11 & 0.22 \\
0.327 & 2.63 & 0.727 & 0.545 & 0.243 & 3.67 & 3.212 & 0.13 & 0.26 & 0.327 & 2.63 & 0.727 & 0.545 & 0.243 & 3.95 & 3.266 & 0.12 & 0.26 \\ \hline
0.371 & 2.89 & 0.706 & 0.561 & 0.199 & 2.02 & 2.473 & 0.07 & 0.18 & 0.371 & 2.89 & 0.706 & 0.561 & 0.199 & 2.26 & 2.582 & 0.075 & 0.2 \\
0.371 & 2.89 & 0.706 & 0.561 & 0.199 & 2.43 & 2.488 & 0.068 & 0.19 & 0.371 & 2.89 & 0.706 & 0.561 & 0.199 & 2.61 & 2.517 & 0.075 & 0.2 \\
0.371 & 2.89 & 0.706 & 0.561 & 0.199 & 2.81 & 2.5 & 0.078 & 0.2 & 0.371 & 2.89 & 0.706 & 0.561 & 0.199 & 3.04 & 2.382 & 0.069 & 0.16 \\
0.371 & 2.89 & 0.706 & 0.561 & 0.199 & 3.28 & 2.306 & 0.067 & 0.16 & 0.371 & 2.89 & 0.706 & 0.561 & 0.199 & 3.51 & 2.309 & 0.064 & 0.19 \\
0.371 & 2.89 & 0.706 & 0.561 & 0.199 & 3.73 & 2.335 & 0.063 & 0.2 & 0.371 & 2.89 & 0.706 & 0.561 & 0.199 & 4.05 & 2.297 & 0.044 & 0.18 \\ \hline
0.163 & 1.49 & 0.826 & 0.48 & 0.427 & 2.52 & 11.15 & 0.48 & 1.1 & 0.163 & 1.49 & 0.826 & 0.48 & 0.427 & 2.71 & 11.62 & 0.52 & 1.1 \\
0.163 & 1.49 & 0.826 & 0.48 & 0.427 & 2.84 & 11.68 & 0.52 & 1.1 & 0.163 & 1.49 & 0.826 & 0.48 & 0.427 & 2.95 & 12.97 & 0.59 & 1.3 \\
0.163 & 1.49 & 0.826 & 0.48 & 0.427 & 3.05 & 10.89 & 0.49 & 1 & 0.163 & 1.49 & 0.826 & 0.48 & 0.427 & 3.15 & 12.46 & 0.65 & 1.2 \\
0.163 & 1.49 & 0.826 & 0.48 & 0.427 & 3.26 & 10.6 & 0.46 & 1 & 0.163 & 1.49 & 0.826 & 0.48 & 0.427 & 3.38 & 11.47 & 0.54 & 1.1 \\
0.163 & 1.49 & 0.826 & 0.48 & 0.427 & 3.59 & 10.47 & 0.33 & 1 & 0.163 & 1.49 & 0.826 & 0.48 & 0.427 & 3.59 & 10.47 & 0.33 & 1 \\ \hline
0.187 & 1.69 & 0.816 & 0.486 & 0.404 & 2.47 & 9.027 & 0.39 & 0.92 & 0.187 & 1.69 & 0.816 & 0.486 & 0.404 & 2.65 & 10.35 & 0.5 & 1.1 \\
0.187 & 1.69 & 0.816 & 0.486 & 0.404 & 2.78 & 9.733 & 0.44 & 1 & 0.187 & 1.69 & 0.816 & 0.486 & 0.404 & 2.9 & 9.54 & 0.44 & 0.99 \\
0.187 & 1.69 & 0.816 & 0.486 & 0.404 & 3.02 & 10.44 & 0.49 & 1.1 & 0.187 & 1.69 & 0.816 & 0.486 & 0.404 & 3.14 & 9.7 & 0.44 & 1 \\
0.187 & 1.69 & 0.816 & 0.486 & 0.404 & 3.26 & 9.801 & 0.44 & 1 & 0.187 & 1.69 & 0.816 & 0.486 & 0.404 & 3.38 & 8.732 & 0.41 & 0.93 \\
0.187 & 1.69 & 0.816 & 0.486 & 0.404 & 3.51 & 8.193 & 0.35 & 0.86 & 0.187 & 1.69 & 0.816 & 0.486 & 0.404 & 3.7 & 10.43 & 0.53 & 1.1 \\ \hline
0.205 & 1.83 & 0.809 & 0.489 & 0.395 & 2.43 & 7.583 & 0.34 & 0.5 & 0.205 & 1.83 & 0.809 & 0.489 & 0.395 & 2.61 & 8.255 & 0.36 & 0.56 \\
0.205 & 1.83 & 0.809 & 0.489 & 0.395 & 2.74 & 7.982 & 0.34 & 0.54 & 0.205 & 1.83 & 0.809 & 0.489 & 0.395 & 2.86 & 7.79 & 0.35 & 0.54 \\
0.205 & 1.83 & 0.809 & 0.489 & 0.395 & 2.99 & 7.596 & 0.32 & 0.52 & 0.205 & 1.83 & 0.809 & 0.489 & 0.395 & 3.12 & 8.067 & 0.38 & 0.55 \\
0.205 & 1.83 & 0.809 & 0.489 & 0.395 & 3.25 & 7.788 & 0.33 & 0.55 & 0.205 & 1.83 & 0.809 & 0.489 & 0.395 & 3.39 & 8.1 & 0.35 & 0.59 \\
0.205 & 1.83 & 0.809 & 0.489 & 0.395 & 3.52 & 7.88 & 0.36 & 0.57 & 0.205 & 1.83 & 0.809 & 0.489 & 0.395 & 3.72 & 7.382 & 0.3 & 0.52 \\ \hline
0.223 & 1.97 & 0.799 & 0.496 & 0.385 & 2.38 & 6.513 & 0.28 & 0.43 & 0.223 & 1.97 & 0.799 & 0.496 & 0.385 & 2.55 & 6.614 & 0.28 & 0.43 \\
0.223 & 1.97 & 0.799 & 0.496 & 0.385 & 2.68 & 6.038 & 0.27 & 0.4 & 0.223 & 1.97 & 0.799 & 0.496 & 0.385 & 2.81 & 6.66 & 0.3 & 0.45 \\
0.223 & 1.97 & 0.799 & 0.496 & 0.385 & 2.96 & 6.588 & 0.3 & 0.44 & 0.223 & 1.97 & 0.799 & 0.496 & 0.385 & 3.12 & 6.341 & 0.27 & 0.39 \\
0.223 & 1.97 & 0.799 & 0.496 & 0.385 & 3.27 & 6.855 & 0.31 & 0.47 & 0.223 & 1.97 & 0.799 & 0.496 & 0.385 & 3.41 & 6.75 & 0.31 & 0.47 \\
0.223 & 1.97 & 0.799 & 0.496 & 0.385 & 3.55 & 6.31 & 0.29 & 0.45 & 0.223 & 1.97 & 0.799 & 0.496 & 0.385 & 3.75 & 6.956 & 0.3 & 0.48 \\ \hline
0.256 & 2.21 & 0.782 & 0.507 & 0.37 & 2.17 & 4.856 & 0.14 & 0.34 & 0.256 & 2.21 & 0.782 & 0.507 & 0.37 & 2.35 & 4.648 & 0.13 & 0.31 \\
0.256 & 2.21 & 0.782 & 0.507 & 0.37 & 2.48 & 5.105 & 0.16 & 0.34 & 0.256 & 2.21 & 0.782 & 0.507 & 0.37 & 2.6 & 5.215 & 0.17 & 0.35 \\
0.256 & 2.21 & 0.782 & 0.507 & 0.37 & 2.74 & 5.025 & 0.16 & 0.34 & 0.256 & 2.21 & 0.782 & 0.507 & 0.37 & 2.93 & 4.618 & 0.15 & 0.32 \\
0.256 & 2.21 & 0.782 & 0.507 & 0.37 & 3.16 & 5.046 & 0.16 & 0.3 & 0.256 & 2.21 & 0.782 & 0.507 & 0.37 & 3.38 & 5.051 & 0.17 & 0.36 \\
0.256 & 2.21 & 0.782 & 0.507 & 0.37 & 3.57 & 4.536 & 0.14 & 0.33 & 0.256 & 2.21 & 0.782 & 0.507 & 0.37 & 3.83 & 4.27 & 0.091 & 0.31 \\ \hline
\end{tabular}
\label{xs_table_2510_m}
\end{table*}

\begin{table*}[ht]
\caption{Unpolarized SIDIS cross section in 2D bins for the $\pi^+$ channel. $Q^2$ is in unit of GeV$^2$, $P_t$ in unit of GeV, $\phi_h$ in unit of rad. $d\sigma$ stands for $d\sigma/(dx_{bj}dydz_{h} d\phi_S dP_t^2d\phi_h)$ in unit of $\rm{nb}\cdot \rm{GeV}^{-2}\cdot rad^{-2}$. Stat. unc. and Sys. unc. stand for statistical and systematic uncertainties, respectively, in unit of $\rm{nb}\cdot \rm{GeV}^{-2}\cdot rad^{-2}$.}
\begin{tabular}{|c c c c c c c c c|c c c c c c c c c|}\hline
$x_{bj}$ & $Q^2$ & y & $z_{h}$ & $P_t$ & $\phi_h$ &  $d\sigma$ & Stat. unc. & Sys. unc. & $x_{bj}$ & $Q^2$ & y & $z_{h}$ & $P_t$ & $\phi_h$ &  $d\sigma$ & Stat. unc. & Sys. unc. \\ \hline
0.164 & 1.5 & 0.825 & 0.48 & 0.36 & 3.11 & 26.34 & 1.5 & 1.6 & 0.164 & 1.5 & 0.825 & 0.48 & 0.378 & 3.11 & 24.07 & 1.3 & 1.6 \\
0.164 & 1.5 & 0.825 & 0.48 & 0.391 & 3.11 & 20.47 & 1 & 1.4 & 0.164 & 1.5 & 0.825 & 0.48 & 0.404 & 3.11 & 19.32 & 0.98 & 1.4 \\
0.164 & 1.5 & 0.825 & 0.48 & 0.416 & 3.11 & 18.93 & 0.96 & 1.3 & 0.164 & 1.5 & 0.825 & 0.48 & 0.428 & 3.11 & 16.92 & 0.86 & 1.2 \\
0.164 & 1.5 & 0.825 & 0.48 & 0.44 & 3.11 & 15.54 & 0.79 & 1.1 & 0.164 & 1.5 & 0.825 & 0.48 & 0.453 & 3.11 & 17.26 & 0.81 & 1.2 \\
0.164 & 1.5 & 0.825 & 0.48 & 0.467 & 3.11 & 14.33 & 0.73 & 1 & 0.164 & 1.5 & 0.825 & 0.48 & 0.49 & 3.11 & 12.23 & 0.52 & 1 \\ \hline
0.19 & 1.71 & 0.814 & 0.487 & 0.335 & 3.08 & 18.79 & 1 & 1 & 0.19 & 1.71 & 0.814 & 0.487 & 0.357 & 3.08 & 19.6 & 1 & 1.2 \\
0.19 & 1.71 & 0.814 & 0.487 & 0.371 & 3.08 & 15.51 & 0.75 & 0.91 & 0.19 & 1.71 & 0.814 & 0.487 & 0.385 & 3.08 & 14.88 & 0.71 & 0.89 \\
0.19 & 1.71 & 0.814 & 0.487 & 0.398 & 3.08 & 15.38 & 0.76 & 0.92 & 0.19 & 1.71 & 0.814 & 0.487 & 0.41 & 3.08 & 16.47 & 0.79 & 1 \\
0.19 & 1.71 & 0.814 & 0.487 & 0.423 & 3.08 & 13.37 & 0.63 & 0.81 & 0.19 & 1.71 & 0.814 & 0.487 & 0.436 & 3.08 & 14.11 & 0.69 & 0.85 \\
0.19 & 1.71 & 0.814 & 0.487 & 0.449 & 3.08 & 14.14 & 0.68 & 0.87 & 0.19 & 1.71 & 0.814 & 0.487 & 0.472 & 3.08 & 11.2 & 0.49 & 0.79 \\ \hline
0.209 & 1.86 & 0.804 & 0.492 & 0.314 & 3.08 & 17.83 & 0.93 & 1 & 0.209 & 1.86 & 0.804 & 0.492 & 0.338 & 3.08 & 16.07 & 0.77 & 0.92 \\
0.209 & 1.86 & 0.804 & 0.492 & 0.353 & 3.08 & 13.83 & 0.68 & 0.8 & 0.209 & 1.86 & 0.804 & 0.492 & 0.366 & 3.08 & 13.82 & 0.68 & 0.8 \\
0.209 & 1.86 & 0.804 & 0.492 & 0.378 & 3.08 & 13.33 & 0.65 & 0.77 & 0.209 & 1.86 & 0.804 & 0.492 & 0.39 & 3.08 & 13.82 & 0.65 & 0.79 \\
0.209 & 1.86 & 0.804 & 0.492 & 0.403 & 3.08 & 11.41 & 0.53 & 0.66 & 0.209 & 1.86 & 0.804 & 0.492 & 0.417 & 3.08 & 11.28 & 0.52 & 0.64 \\
0.209 & 1.86 & 0.804 & 0.492 & 0.432 & 3.08 & 11.48 & 0.53 & 0.67 & 0.209 & 1.86 & 0.804 & 0.492 & 0.458 & 3.08 & 10.01 & 0.42 & 0.65 \\ \hline
0.227 & 1.99 & 0.794 & 0.499 & 0.291 & 3.08 & 14.96 & 0.79 & 0.91 & 0.227 & 1.99 & 0.794 & 0.499 & 0.316 & 3.08 & 13.16 & 0.69 & 0.74 \\
0.227 & 1.99 & 0.794 & 0.499 & 0.332 & 3.08 & 11.94 & 0.58 & 0.67 & 0.227 & 1.99 & 0.794 & 0.499 & 0.347 & 3.08 & 10.47 & 0.51 & 0.57 \\
0.227 & 1.99 & 0.794 & 0.499 & 0.36 & 3.08 & 11.48 & 0.56 & 0.68 & 0.227 & 1.99 & 0.794 & 0.499 & 0.372 & 3.08 & 11.85 & 0.56 & 0.65 \\
0.227 & 1.99 & 0.794 & 0.499 & 0.384 & 3.08 & 10.77 & 0.52 & 0.6 & 0.227 & 1.99 & 0.794 & 0.499 & 0.397 & 3.08 & 10.79 & 0.49 & 0.61 \\
0.227 & 1.99 & 0.794 & 0.499 & 0.412 & 3.08 & 9.563 & 0.46 & 0.54 & 0.227 & 1.99 & 0.794 & 0.499 & 0.44 & 3.08 & 8.39 & 0.34 & 0.5 \\ \hline
0.245 & 2.11 & 0.781 & 0.507 & 0.271 & 3.06 & 13.5 & 0.73 & 0.78 & 0.245 & 2.11 & 0.781 & 0.507 & 0.297 & 3.06 & 10.26 & 0.53 & 0.6 \\
0.245 & 2.11 & 0.781 & 0.507 & 0.314 & 3.06 & 10.58 & 0.51 & 0.6 & 0.245 & 2.11 & 0.781 & 0.507 & 0.328 & 3.06 & 11.52 & 0.58 & 0.68 \\
0.245 & 2.11 & 0.781 & 0.507 & 0.34 & 3.06 & 10.2 & 0.52 & 0.53 & 0.245 & 2.11 & 0.781 & 0.507 & 0.352 & 3.06 & 10.48 & 0.52 & 0.57 \\
0.245 & 2.11 & 0.781 & 0.507 & 0.365 & 3.06 & 8.899 & 0.42 & 0.5 & 0.245 & 2.11 & 0.781 & 0.507 & 0.379 & 3.06 & 10.52 & 0.49 & 0.58 \\
0.245 & 2.11 & 0.781 & 0.507 & 0.394 & 3.06 & 8.806 & 0.42 & 0.49 & 0.245 & 2.11 & 0.781 & 0.507 & 0.425 & 3.06 & 8.986 & 0.37 & 0.52 \\ \hline
0.264 & 2.24 & 0.769 & 0.515 & 0.247 & 3.09 & 10.62 & 0.57 & 0.66 & 0.264 & 2.24 & 0.769 & 0.515 & 0.274 & 3.09 & 9.329 & 0.46 & 0.58 \\
0.264 & 2.24 & 0.769 & 0.515 & 0.292 & 3.09 & 8.852 & 0.43 & 0.53 & 0.264 & 2.24 & 0.769 & 0.515 & 0.307 & 3.09 & 8.249 & 0.4 & 0.48 \\
0.264 & 2.24 & 0.769 & 0.515 & 0.321 & 3.09 & 8.493 & 0.39 & 0.49 & 0.264 & 2.24 & 0.769 & 0.515 & 0.334 & 3.09 & 8.379 & 0.41 & 0.51 \\
0.264 & 2.24 & 0.769 & 0.515 & 0.346 & 3.09 & 8.113 & 0.4 & 0.48 & 0.264 & 2.24 & 0.769 & 0.515 & 0.359 & 3.09 & 7.779 & 0.37 & 0.45 \\
0.264 & 2.24 & 0.769 & 0.515 & 0.375 & 3.09 & 8.01 & 0.37 & 0.47 & 0.264 & 2.24 & 0.769 & 0.515 & 0.407 & 3.09 & 6.983 & 0.28 & 0.4 \\ \hline
0.285 & 2.37 & 0.755 & 0.524 & 0.221 & 3.07 & 9.644 & 0.5 & 0.61 & 0.285 & 2.37 & 0.755 & 0.524 & 0.25 & 3.07 & 8.818 & 0.42 & 0.6 \\
0.285 & 2.37 & 0.755 & 0.524 & 0.268 & 3.07 & 7.823 & 0.38 & 0.48 & 0.285 & 2.37 & 0.755 & 0.524 & 0.283 & 3.07 & 7.046 & 0.36 & 0.43 \\
0.285 & 2.37 & 0.755 & 0.524 & 0.297 & 3.07 & 7.482 & 0.34 & 0.47 & 0.285 & 2.37 & 0.755 & 0.524 & 0.31 & 3.07 & 7.382 & 0.36 & 0.43 \\
0.285 & 2.37 & 0.755 & 0.524 & 0.323 & 3.07 & 6.446 & 0.3 & 0.38 & 0.285 & 2.37 & 0.755 & 0.524 & 0.337 & 3.07 & 7.356 & 0.35 & 0.46 \\
0.285 & 2.37 & 0.755 & 0.524 & 0.352 & 3.07 & 6.807 & 0.34 & 0.42 & 0.285 & 2.37 & 0.755 & 0.524 & 0.385 & 3.07 & 6.245 & 0.24 & 0.38 \\ \hline
0.308 & 2.52 & 0.739 & 0.536 & 0.193 & 3.07 & 8.177 & 0.39 & 0.5 & 0.308 & 2.52 & 0.739 & 0.536 & 0.223 & 3.07 & 7.161 & 0.34 & 0.46 \\
0.308 & 2.52 & 0.739 & 0.536 & 0.242 & 3.07 & 7.528 & 0.34 & 0.46 & 0.308 & 2.52 & 0.739 & 0.536 & 0.257 & 3.07 & 6.659 & 0.31 & 0.42 \\
0.308 & 2.52 & 0.739 & 0.536 & 0.271 & 3.07 & 6.947 & 0.31 & 0.46 & 0.308 & 2.52 & 0.739 & 0.536 & 0.285 & 3.07 & 6.056 & 0.27 & 0.38 \\
0.308 & 2.52 & 0.739 & 0.536 & 0.299 & 3.07 & 5.84 & 0.27 & 0.37 & 0.308 & 2.52 & 0.739 & 0.536 & 0.314 & 3.07 & 5.786 & 0.26 & 0.38 \\
0.308 & 2.52 & 0.739 & 0.536 & 0.331 & 3.07 & 5.641 & 0.25 & 0.34 & 0.308 & 2.52 & 0.739 & 0.536 & 0.368 & 3.07 & 5.696 & 0.22 & 0.35 \\ \hline
0.336 & 2.68 & 0.722 & 0.549 & 0.161 & 3.1 & 5.965 & 0.29 & 0.38 & 0.336 & 2.68 & 0.722 & 0.549 & 0.195 & 3.1 & 5.128 & 0.23 & 0.34 \\
0.336 & 2.68 & 0.722 & 0.549 & 0.214 & 3.1 & 5.22 & 0.25 & 0.33 & 0.336 & 2.68 & 0.722 & 0.549 & 0.229 & 3.1 & 5.272 & 0.25 & 0.33 \\
0.336 & 2.68 & 0.722 & 0.549 & 0.243 & 3.1 & 5.021 & 0.23 & 0.32 & 0.336 & 2.68 & 0.722 & 0.549 & 0.257 & 3.1 & 5.141 & 0.24 & 0.33 \\
0.336 & 2.68 & 0.722 & 0.549 & 0.271 & 3.1 & 5.085 & 0.23 & 0.35 & 0.336 & 2.68 & 0.722 & 0.549 & 0.286 & 3.1 & 5.428 & 0.24 & 0.36 \\
0.336 & 2.68 & 0.722 & 0.549 & 0.304 & 3.1 & 5.161 & 0.23 & 0.33 & 0.336 & 2.68 & 0.722 & 0.549 & 0.343 & 3.1 & 4.42 & 0.16 & 0.28 \\ \hline
0.381 & 2.96 & 0.704 & 0.562 & 0.122 & 3.05 & 4.665 & 0.19 & 0.35 & 0.381 & 2.96 & 0.704 & 0.562 & 0.155 & 3.05 & 4.346 & 0.17 & 0.36 \\
0.381 & 2.96 & 0.704 & 0.562 & 0.175 & 3.05 & 4.128 & 0.16 & 0.33 & 0.381 & 2.96 & 0.704 & 0.562 & 0.192 & 3.05 & 4.356 & 0.16 & 0.37 \\
0.381 & 2.96 & 0.704 & 0.562 & 0.207 & 3.05 & 4.13 & 0.16 & 0.34 & 0.381 & 2.96 & 0.704 & 0.562 & 0.221 & 3.05 & 4.455 & 0.16 & 0.37 \\
0.381 & 2.96 & 0.704 & 0.562 & 0.236 & 3.05 & 3.989 & 0.15 & 0.35 & 0.381 & 2.96 & 0.704 & 0.562 & 0.253 & 3.05 & 3.558 & 0.13 & 0.31 \\
0.381 & 2.96 & 0.704 & 0.562 & 0.272 & 3.05 & 3.516 & 0.13 & 0.31 & 0.381 & 2.96 & 0.704 & 0.562 & 0.316 & 3.05 & 3.496 & 0.11 & 0.33 \\ \hline
\end{tabular}
\label{xs_table_1010_p}
\end{table*}

\begin{table*}[ht]
\caption{Unpolarized SIDIS cross section in 2D bins for the $\pi^-$ channel. $Q^2$ is in unit of GeV$^2$, $P_t$ in unit of GeV, $\phi_h$ in unit of rad. $d\sigma$ stands for $d\sigma/(dx_{bj}dydz_{h} d\phi_S dP_t^2d\phi_h)$ in unit of $\rm{nb}\cdot \rm{GeV}^{-2}\cdot rad^{-2}$. Stat. unc. and Sys. unc. stand for statistical and systematic uncertainties, respectively, in unit of $\rm{nb}\cdot \rm{GeV}^{-2}\cdot rad^{-2}$.}
\begin{tabular}{|c c c c c c c c c|c c c c c c c c c|}\hline
$x_{bj}$ & $Q^2$ & y & $z_{h}$ & $P_t$ & $\phi_h$ &  $d\sigma$ & Stat. unc. & Sys. unc. & $x_{bj}$ & $Q^2$ & y & $z_{h}$ & $P_t$ & $\phi_h$ &  $d\sigma$ & Stat. unc. & Sys. unc. \\ \hline
0.164 & 1.5 & 0.825 & 0.48 & 0.36 & 3.11 & 18.42 & 0.99 & 1.2 & 0.164 & 1.5 & 0.825 & 0.48 & 0.378 & 3.11 & 16.67 & 0.86 & 1.2 \\
0.164 & 1.5 & 0.825 & 0.48 & 0.391 & 3.11 & 13.39 & 0.59 & 1 & 0.164 & 1.5 & 0.825 & 0.48 & 0.404 & 3.11 & 11.7 & 0.51 & 0.89 \\
0.164 & 1.5 & 0.825 & 0.48 & 0.416 & 3.11 & 10.98 & 0.46 & 0.83 & 0.164 & 1.5 & 0.825 & 0.48 & 0.428 & 3.11 & 10.04 & 0.43 & 0.75 \\
0.164 & 1.5 & 0.825 & 0.48 & 0.44 & 3.11 & 9.779 & 0.42 & 0.73 & 0.164 & 1.5 & 0.825 & 0.48 & 0.453 & 3.11 & 9.9 & 0.38 & 0.72 \\
0.164 & 1.5 & 0.825 & 0.48 & 0.467 & 3.11 & 9.662 & 0.44 & 0.74 & 0.164 & 1.5 & 0.825 & 0.48 & 0.49 & 3.11 & 8.378 & 0.31 & 0.7 \\ \hline
0.19 & 1.71 & 0.814 & 0.487 & 0.335 & 3.08 & 13.14 & 0.65 & 0.82 & 0.19 & 1.71 & 0.814 & 0.487 & 0.357 & 3.08 & 12.05 & 0.54 & 0.8 \\
0.19 & 1.71 & 0.814 & 0.487 & 0.371 & 3.08 & 10.79 & 0.48 & 0.71 & 0.19 & 1.71 & 0.814 & 0.487 & 0.385 & 3.08 & 9.442 & 0.39 & 0.64 \\
0.19 & 1.71 & 0.814 & 0.487 & 0.398 & 3.08 & 10.11 & 0.45 & 0.68 & 0.19 & 1.71 & 0.814 & 0.487 & 0.41 & 3.08 & 8.966 & 0.35 & 0.6 \\
0.19 & 1.71 & 0.814 & 0.487 & 0.423 & 3.08 & 8.546 & 0.35 & 0.57 & 0.19 & 1.71 & 0.814 & 0.487 & 0.436 & 3.08 & 8.429 & 0.35 & 0.56 \\
0.19 & 1.71 & 0.814 & 0.487 & 0.449 & 3.08 & 7.746 & 0.3 & 0.53 & 0.19 & 1.71 & 0.814 & 0.487 & 0.472 & 3.08 & 7.314 & 0.28 & 0.54 \\ \hline
0.209 & 1.86 & 0.804 & 0.491 & 0.314 & 3.08 & 10.65 & 0.49 & 0.67 & 0.209 & 1.86 & 0.804 & 0.491 & 0.338 & 3.08 & 8.655 & 0.35 & 0.56 \\
0.209 & 1.86 & 0.804 & 0.491 & 0.353 & 3.08 & 8.671 & 0.38 & 0.57 & 0.209 & 1.86 & 0.804 & 0.491 & 0.366 & 3.08 & 7.718 & 0.32 & 0.51 \\
0.209 & 1.86 & 0.804 & 0.491 & 0.378 & 3.08 & 8.551 & 0.37 & 0.57 & 0.209 & 1.86 & 0.804 & 0.491 & 0.39 & 3.08 & 8.066 & 0.32 & 0.52 \\
0.209 & 1.86 & 0.804 & 0.491 & 0.403 & 3.08 & 7.268 & 0.29 & 0.48 & 0.209 & 1.86 & 0.804 & 0.491 & 0.417 & 3.08 & 6.871 & 0.27 & 0.45 \\
0.209 & 1.86 & 0.804 & 0.491 & 0.432 & 3.08 & 6.267 & 0.23 & 0.42 & 0.209 & 1.86 & 0.804 & 0.491 & 0.458 & 3.08 & 5.94 & 0.2 & 0.42 \\ \hline
0.227 & 1.99 & 0.794 & 0.498 & 0.291 & 3.08 & 8.909 & 0.41 & 0.56 & 0.227 & 1.99 & 0.794 & 0.498 & 0.316 & 3.08 & 9.172 & 0.45 & 0.59 \\
0.227 & 1.99 & 0.794 & 0.498 & 0.332 & 3.08 & 7.514 & 0.32 & 0.49 & 0.227 & 1.99 & 0.794 & 0.498 & 0.347 & 3.08 & 7.359 & 0.33 & 0.48 \\
0.227 & 1.99 & 0.794 & 0.498 & 0.36 & 3.08 & 6.392 & 0.26 & 0.41 & 0.227 & 1.99 & 0.794 & 0.498 & 0.372 & 3.08 & 6.896 & 0.28 & 0.44 \\
0.227 & 1.99 & 0.794 & 0.498 & 0.384 & 3.08 & 7.082 & 0.31 & 0.46 & 0.227 & 1.99 & 0.794 & 0.498 & 0.397 & 3.08 & 5.912 & 0.22 & 0.39 \\
0.227 & 1.99 & 0.794 & 0.498 & 0.412 & 3.08 & 6.607 & 0.29 & 0.42 & 0.227 & 1.99 & 0.794 & 0.498 & 0.44 & 3.08 & 4.884 & 0.16 & 0.32 \\ \hline
0.245 & 2.11 & 0.781 & 0.507 & 0.271 & 3.06 & 8.322 & 0.4 & 0.53 & 0.245 & 2.11 & 0.781 & 0.507 & 0.297 & 3.06 & 6.505 & 0.3 & 0.44 \\
0.245 & 2.11 & 0.781 & 0.507 & 0.314 & 3.06 & 6.682 & 0.29 & 0.44 & 0.245 & 2.11 & 0.781 & 0.507 & 0.328 & 3.06 & 6.776 & 0.3 & 0.45 \\
0.245 & 2.11 & 0.781 & 0.507 & 0.34 & 3.06 & 6.682 & 0.31 & 0.43 & 0.245 & 2.11 & 0.781 & 0.507 & 0.352 & 3.06 & 6.197 & 0.26 & 0.41 \\
0.245 & 2.11 & 0.781 & 0.507 & 0.365 & 3.06 & 5.828 & 0.25 & 0.38 & 0.245 & 2.11 & 0.781 & 0.507 & 0.379 & 3.06 & 5.559 & 0.21 & 0.36 \\
0.245 & 2.11 & 0.781 & 0.507 & 0.394 & 3.06 & 5.139 & 0.2 & 0.33 & 0.245 & 2.11 & 0.781 & 0.507 & 0.425 & 3.06 & 4.673 & 0.15 & 0.3 \\ \hline
0.264 & 2.24 & 0.769 & 0.515 & 0.247 & 3.09 & 7.03 & 0.34 & 0.47 & 0.264 & 2.24 & 0.769 & 0.515 & 0.274 & 3.09 & 5.788 & 0.25 & 0.39 \\
0.264 & 2.24 & 0.769 & 0.515 & 0.292 & 3.09 & 5.678 & 0.25 & 0.38 & 0.264 & 2.24 & 0.769 & 0.515 & 0.307 & 3.09 & 5.4 & 0.23 & 0.37 \\
0.264 & 2.24 & 0.769 & 0.515 & 0.321 & 3.09 & 5.122 & 0.2 & 0.34 & 0.264 & 2.24 & 0.769 & 0.515 & 0.334 & 3.09 & 4.738 & 0.2 & 0.32 \\
0.264 & 2.24 & 0.769 & 0.515 & 0.346 & 3.09 & 5.25 & 0.23 & 0.36 & 0.264 & 2.24 & 0.769 & 0.515 & 0.359 & 3.09 & 4.581 & 0.19 & 0.3 \\
0.264 & 2.24 & 0.769 & 0.515 & 0.375 & 3.09 & 4.161 & 0.15 & 0.28 & 0.264 & 2.24 & 0.769 & 0.515 & 0.407 & 3.09 & 3.952 & 0.13 & 0.26 \\ \hline
0.284 & 2.37 & 0.755 & 0.524 & 0.221 & 3.07 & 5.468 & 0.25 & 0.36 & 0.284 & 2.37 & 0.755 & 0.524 & 0.25 & 3.07 & 4.766 & 0.19 & 0.33 \\
0.284 & 2.37 & 0.755 & 0.524 & 0.268 & 3.07 & 4.425 & 0.18 & 0.31 & 0.284 & 2.37 & 0.755 & 0.524 & 0.283 & 3.07 & 4.722 & 0.21 & 0.33 \\
0.284 & 2.37 & 0.755 & 0.524 & 0.297 & 3.07 & 3.927 & 0.15 & 0.27 & 0.284 & 2.37 & 0.755 & 0.524 & 0.31 & 3.07 & 4.239 & 0.17 & 0.29 \\
0.284 & 2.37 & 0.755 & 0.524 & 0.323 & 3.07 & 4.3 & 0.18 & 0.3 & 0.284 & 2.37 & 0.755 & 0.524 & 0.337 & 3.07 & 3.641 & 0.14 & 0.25 \\
0.284 & 2.37 & 0.755 & 0.524 & 0.352 & 3.07 & 4.377 & 0.19 & 0.3 & 0.284 & 2.37 & 0.755 & 0.524 & 0.385 & 3.07 & 3.491 & 0.11 & 0.24 \\ \hline
0.308 & 2.52 & 0.74 & 0.536 & 0.193 & 3.07 & 4.535 & 0.18 & 0.31 & 0.308 & 2.52 & 0.74 & 0.536 & 0.223 & 3.07 & 4.266 & 0.18 & 0.3 \\
0.308 & 2.52 & 0.74 & 0.536 & 0.242 & 3.07 & 4.076 & 0.16 & 0.29 & 0.308 & 2.52 & 0.74 & 0.536 & 0.257 & 3.07 & 3.858 & 0.15 & 0.28 \\
0.308 & 2.52 & 0.74 & 0.536 & 0.271 & 3.07 & 3.519 & 0.13 & 0.25 & 0.308 & 2.52 & 0.74 & 0.536 & 0.285 & 3.07 & 3.58 & 0.14 & 0.26 \\
0.308 & 2.52 & 0.74 & 0.536 & 0.299 & 3.07 & 3.271 & 0.12 & 0.23 & 0.308 & 2.52 & 0.74 & 0.536 & 0.314 & 3.07 & 3.046 & 0.11 & 0.22 \\
0.308 & 2.52 & 0.74 & 0.536 & 0.331 & 3.07 & 2.705 & 0.094 & 0.19 & 0.308 & 2.52 & 0.74 & 0.536 & 0.368 & 3.07 & 2.972 & 0.091 & 0.21 \\ \hline
0.336 & 2.68 & 0.722 & 0.548 & 0.161 & 3.1 & 3.508 & 0.14 & 0.24 & 0.336 & 2.68 & 0.722 & 0.548 & 0.195 & 3.1 & 2.964 & 0.11 & 0.21 \\
0.336 & 2.68 & 0.722 & 0.548 & 0.214 & 3.1 & 3.319 & 0.14 & 0.24 & 0.336 & 2.68 & 0.722 & 0.548 & 0.229 & 3.1 & 3.235 & 0.14 & 0.23 \\
0.336 & 2.68 & 0.722 & 0.548 & 0.243 & 3.1 & 2.994 & 0.12 & 0.22 & 0.336 & 2.68 & 0.722 & 0.548 & 0.257 & 3.1 & 2.565 & 0.094 & 0.19 \\
0.336 & 2.68 & 0.722 & 0.548 & 0.271 & 3.1 & 2.585 & 0.096 & 0.19 & 0.336 & 2.68 & 0.722 & 0.548 & 0.286 & 3.1 & 2.293 & 0.074 & 0.17 \\
0.336 & 2.68 & 0.722 & 0.548 & 0.304 & 3.1 & 2.579 & 0.09 & 0.19 & 0.336 & 2.68 & 0.722 & 0.548 & 0.343 & 3.1 & 2.416 & 0.072 & 0.17 \\ \hline
0.38 & 2.96 & 0.705 & 0.561 & 0.122 & 3.05 & 3.01 & 0.11 & 0.27 & 0.38 & 2.96 & 0.705 & 0.561 & 0.155 & 3.05 & 2.238 & 0.068 & 0.21 \\
0.38 & 2.96 & 0.705 & 0.561 & 0.175 & 3.05 & 2.398 & 0.077 & 0.24 & 0.38 & 2.96 & 0.705 & 0.561 & 0.192 & 3.05 & 2.295 & 0.068 & 0.23 \\
0.38 & 2.96 & 0.705 & 0.561 & 0.207 & 3.05 & 2.321 & 0.074 & 0.23 & 0.38 & 2.96 & 0.705 & 0.561 & 0.221 & 3.05 & 2.332 & 0.069 & 0.24 \\
0.38 & 2.96 & 0.705 & 0.561 & 0.236 & 3.05 & 1.915 & 0.053 & 0.2 & 0.38 & 2.96 & 0.705 & 0.561 & 0.253 & 3.05 & 2.003 & 0.061 & 0.21 \\
0.38 & 2.96 & 0.705 & 0.561 & 0.272 & 3.05 & 1.889 & 0.058 & 0.2 & 0.38 & 2.96 & 0.705 & 0.561 & 0.316 & 3.05 & 1.713 & 0.039 & 0.2 \\ \hline
\end{tabular}
\label{xs_table_1010_m}
\end{table*}

\clearpage

\nocite{*}
\bibliography{references}

\begin{thebibliography}{46}%
\makeatletter
\providecommand \@ifxundefined [1]{%
 \@ifx{#1\undefined}
}%
\providecommand \@ifnum [1]{%
 \ifnum #1\expandafter \@firstoftwo
 \else \expandafter \@secondoftwo
 \fi
}%
\providecommand \@ifx [1]{%
 \ifx #1\expandafter \@firstoftwo
 \else \expandafter \@secondoftwo
 \fi
}%
\providecommand \natexlab [1]{#1}%
\providecommand \enquote  [1]{``#1''}%
\providecommand \bibnamefont  [1]{#1}%
\providecommand \bibfnamefont [1]{#1}%
\providecommand \citenamefont [1]{#1}%
\providecommand \href@noop [0]{\@secondoftwo}%
\providecommand \href [0]{\begingroup \@sanitize@url \@href}%
\providecommand \@href[1]{\@@startlink{#1}\@@href}%
\providecommand \@@href[1]{\endgroup#1\@@endlink}%
\providecommand \@sanitize@url [0]{\catcode `\\12\catcode `\$12\catcode
  `\&12\catcode `\#12\catcode `\^12\catcode `\_12\catcode `\%12\relax}%
\providecommand \@@startlink[1]{}%
\providecommand \@@endlink[0]{}%
\providecommand \url  [0]{\begingroup\@sanitize@url \@url }%
\providecommand \@url [1]{\endgroup\@href {#1}{\urlprefix }}%
\providecommand \urlprefix  [0]{URL }%
\providecommand \Eprint [0]{\href }%
\providecommand \doibase [0]{http://dx.doi.org/}%
\providecommand \selectlanguage [0]{\@gobble}%
\providecommand \bibinfo  [0]{\@secondoftwo}%
\providecommand \bibfield  [0]{\@secondoftwo}%
\providecommand \translation [1]{[#1]}%
\providecommand \BibitemOpen [0]{}%
\providecommand \bibitemStop [0]{}%
\providecommand \bibitemNoStop [0]{.\EOS\space}%
\providecommand \EOS [0]{\spacefactor3000\relax}%
\providecommand \BibitemShut  [1]{\csname bibitem#1\endcsname}%
\let\auto@bib@innerbib\@empty
\bibitem [{\citenamefont {Arneodo}\ \emph {et~al.}(1987)\citenamefont {Arneodo}
  \emph {et~al.}}]{EMC_mod}%
  \BibitemOpen
  \bibfield  {author} {\bibinfo {author} {\bibfnamefont {M.}~\bibnamefont
  {Arneodo}} \emph {et~al.} (\bibinfo {collaboration} {European Muon}),\ }\href
  {\doibase 10.1007/BF01548808} {\bibfield  {journal} {\bibinfo  {journal} {Z.
  Phys. C}\ }\textbf {\bibinfo {volume} {34}},\ \bibinfo {pages} {277}
  (\bibinfo {year} {1987})}\BibitemShut {NoStop}%
\bibitem [{\citenamefont {Airapetian}\ \emph {et~al.}(2005)\citenamefont
  {Airapetian} \emph {et~al.}}]{HERMES_0}%
  \BibitemOpen
  \bibfield  {author} {\bibinfo {author} {\bibfnamefont {A.}~\bibnamefont
  {Airapetian}} \emph {et~al.} (\bibinfo {collaboration} {HERMES}),\ }\href
  {\doibase 10.1103/PhysRevLett.94.012002} {\bibfield  {journal} {\bibinfo
  {journal} {Phys. Rev. Lett.}\ }\textbf {\bibinfo {volume} {94}},\ \bibinfo
  {pages} {012002} (\bibinfo {year} {2005})}\BibitemShut {NoStop}%
\bibitem [{\citenamefont {Alexakhin}\ \emph {et~al.}(2005)\citenamefont
  {Alexakhin} \emph {et~al.}}]{COMPASS_0}%
  \BibitemOpen
  \bibfield  {author} {\bibinfo {author} {\bibfnamefont {V.~{\relax Yu}.}\
  \bibnamefont {Alexakhin}} \emph {et~al.} (\bibinfo {collaboration}
  {COMPASS}),\ }\href {\doibase 10.1103/PhysRevLett.94.202002} {\bibfield
  {journal} {\bibinfo  {journal} {Phys. Rev. Lett.}\ }\textbf {\bibinfo
  {volume} {94}},\ \bibinfo {pages} {202002} (\bibinfo {year}
  {2005})}\BibitemShut {NoStop}%
\bibitem [{\citenamefont {Avakian}\ \emph {et~al.}(2004)\citenamefont {Avakian}
  \emph {et~al.}}]{CLAS_0}%
  \BibitemOpen
  \bibfield  {author} {\bibinfo {author} {\bibfnamefont {H.}~\bibnamefont
  {Avakian}} \emph {et~al.} (\bibinfo {collaboration} {CLAS}),\ }\href
  {\doibase 10.1103/PhysRevD.69.112004} {\bibfield  {journal} {\bibinfo
  {journal} {Phys. Rev. D}\ }\textbf {\bibinfo {volume} {69}},\ \bibinfo
  {pages} {112004} (\bibinfo {year} {2004})}\BibitemShut {NoStop}%
\bibitem [{\citenamefont {Anselmino}\ \emph {et~al.}(2005)\citenamefont
  {Anselmino}, \citenamefont {Boglione}, \citenamefont {D'Alesio},
  \citenamefont {Kotzinian}, \citenamefont {Murgia},\ and\ \citenamefont
  {Prokudin}}]{Anselmino_2005}%
  \BibitemOpen
  \bibfield  {author} {\bibinfo {author} {\bibfnamefont {M.}~\bibnamefont
  {Anselmino}}, \bibinfo {author} {\bibfnamefont {M.}~\bibnamefont {Boglione}},
  \bibinfo {author} {\bibfnamefont {U.}~\bibnamefont {D'Alesio}}, \bibinfo
  {author} {\bibfnamefont {A.}~\bibnamefont {Kotzinian}}, \bibinfo {author}
  {\bibfnamefont {F.}~\bibnamefont {Murgia}}, \ and\ \bibinfo {author}
  {\bibfnamefont {A.}~\bibnamefont {Prokudin}},\ }\href {\doibase
  10.1103/PhysRevD.71.074006} {\bibfield  {journal} {\bibinfo  {journal} {Phys.
  Rev. D}\ }\textbf {\bibinfo {volume} {71}},\ \bibinfo {pages} {074006}
  (\bibinfo {year} {2005})}\BibitemShut {NoStop}%
\bibitem [{\citenamefont {Anselmino}\ \emph {et~al.}(2006)\citenamefont
  {Anselmino}, \citenamefont {Efremov}, \citenamefont {Kotzinian},\ and\
  \citenamefont {Parsamyan}}]{Anselmino_0}%
  \BibitemOpen
  \bibfield  {author} {\bibinfo {author} {\bibfnamefont {M.}~\bibnamefont
  {Anselmino}}, \bibinfo {author} {\bibfnamefont {A.}~\bibnamefont {Efremov}},
  \bibinfo {author} {\bibfnamefont {A.}~\bibnamefont {Kotzinian}}, \ and\
  \bibinfo {author} {\bibfnamefont {B.}~\bibnamefont {Parsamyan}},\ }\href
  {\doibase 10.1103/PhysRevD.74.074015} {\bibfield  {journal} {\bibinfo
  {journal} {Phys. Rev. D}\ }\textbf {\bibinfo {volume} {74}},\ \bibinfo
  {pages} {074015} (\bibinfo {year} {2006})}\BibitemShut {NoStop}%
\bibitem [{\citenamefont {Mulders}\ and\ \citenamefont
  {Tangerman}(1996)}]{Mulders_TMD}%
  \BibitemOpen
  \bibfield  {author} {\bibinfo {author} {\bibfnamefont {P.~J.}\ \bibnamefont
  {Mulders}}\ and\ \bibinfo {author} {\bibfnamefont {R.~D.}\ \bibnamefont
  {Tangerman}},\ }\href {\doibase 10.1016/S0550-3213(96)00648-7,
  10.1016/0550-3213(95)00632-X} {\bibfield  {journal} {\bibinfo  {journal}
  {Nucl. Phys.}\ }\textbf {\bibinfo {volume} {B461}},\ \bibinfo {pages} {197}
  (\bibinfo {year} {1996})},\ \bibinfo {note} {[Erratum: Nucl.
  Phys.B484,538(1997)]}\BibitemShut {NoStop}%
\bibitem [{\citenamefont {Boer}\ and\ \citenamefont
  {Mulders}(1998)}]{BM_original}%
  \BibitemOpen
  \bibfield  {author} {\bibinfo {author} {\bibfnamefont {D.}~\bibnamefont
  {Boer}}\ and\ \bibinfo {author} {\bibfnamefont {P.~J.}\ \bibnamefont
  {Mulders}},\ }\href {\doibase 10.1103/PhysRevD.57.5780} {\bibfield  {journal}
  {\bibinfo  {journal} {Phys. Rev. D}\ }\textbf {\bibinfo {volume} {57}},\
  \bibinfo {pages} {5780} (\bibinfo {year} {1998})}\BibitemShut {NoStop}%
\bibitem [{\citenamefont {Collins}\ \emph {et~al.}(1985)\citenamefont
  {Collins}, \citenamefont {Soper},\ and\ \citenamefont
  {Sterman}}]{Collins_tmd_fac}%
  \BibitemOpen
  \bibfield  {author} {\bibinfo {author} {\bibfnamefont {J.~C.}\ \bibnamefont
  {Collins}}, \bibinfo {author} {\bibfnamefont {D.~E.}\ \bibnamefont {Soper}},
  \ and\ \bibinfo {author} {\bibfnamefont {G.~F.}\ \bibnamefont {Sterman}},\
  }\href {\doibase 10.1016/0550-3213(85)90479-1} {\bibfield  {journal}
  {\bibinfo  {journal} {Nucl. Phys. B}\ }\textbf {\bibinfo {volume} {250}},\
  \bibinfo {pages} {199} (\bibinfo {year} {1985})}\BibitemShut {NoStop}%
\bibitem [{\citenamefont {Brodsky}\ \emph {et~al.}(2002)\citenamefont
  {Brodsky}, \citenamefont {Hwang},\ and\ \citenamefont
  {Schmidt}}]{Brodsky_tmd_fac}%
  \BibitemOpen
  \bibfield  {author} {\bibinfo {author} {\bibfnamefont {S.~J.}\ \bibnamefont
  {Brodsky}}, \bibinfo {author} {\bibfnamefont {D.~S.}\ \bibnamefont {Hwang}},
  \ and\ \bibinfo {author} {\bibfnamefont {I.}~\bibnamefont {Schmidt}},\ }\href
  {\doibase 10.1016/S0370-2693(02)01320-5} {\bibfield  {journal} {\bibinfo
  {journal} {Phys. Lett.}\ }\textbf {\bibinfo {volume} {B530}},\ \bibinfo
  {pages} {99} (\bibinfo {year} {2002})}\BibitemShut {NoStop}%
\bibitem [{\citenamefont {Ji}\ \emph {et~al.}(2004)\citenamefont {Ji},
  \citenamefont {Ma},\ and\ \citenamefont {Yuan}}]{Ji_tmd_fac}%
  \BibitemOpen
  \bibfield  {author} {\bibinfo {author} {\bibfnamefont {X.-d.}\ \bibnamefont
  {Ji}}, \bibinfo {author} {\bibfnamefont {J.-P.}\ \bibnamefont {Ma}}, \ and\
  \bibinfo {author} {\bibfnamefont {F.}~\bibnamefont {Yuan}},\ }\href {\doibase
  10.1016/j.physletb.2004.07.026} {\bibfield  {journal} {\bibinfo  {journal}
  {Phys. Lett. B}\ }\textbf {\bibinfo {volume} {597}},\ \bibinfo {pages} {299}
  (\bibinfo {year} {2004})}\BibitemShut {NoStop}%
\bibitem [{\citenamefont {Aybat}\ and\ \citenamefont
  {Rogers}(2011)}]{Aybat_review}%
  \BibitemOpen
  \bibfield  {author} {\bibinfo {author} {\bibfnamefont {S.~M.}\ \bibnamefont
  {Aybat}}\ and\ \bibinfo {author} {\bibfnamefont {T.~C.}\ \bibnamefont
  {Rogers}},\ }\href {\doibase 10.1103/PhysRevD.83.114042} {\bibfield
  {journal} {\bibinfo  {journal} {Phys. Rev. D}\ }\textbf {\bibinfo {volume}
  {83}},\ \bibinfo {pages} {114042} (\bibinfo {year} {2011})}\BibitemShut
  {NoStop}%
\bibitem [{\citenamefont {Bacchetta}\ \emph {et~al.}()\citenamefont
  {Bacchetta}, \citenamefont {Diehl}, \citenamefont {Goeke}, \citenamefont
  {Metz}, \citenamefont {Mulders},\ and\ \citenamefont
  {Schlegel}}]{Bacchetta_basis}%
  \BibitemOpen
  \bibfield  {author} {\bibinfo {author} {\bibfnamefont {A.}~\bibnamefont
  {Bacchetta}}, \bibinfo {author} {\bibfnamefont {M.}~\bibnamefont {Diehl}},
  \bibinfo {author} {\bibfnamefont {K.}~\bibnamefont {Goeke}}, \bibinfo
  {author} {\bibfnamefont {A.}~\bibnamefont {Metz}}, \bibinfo {author}
  {\bibfnamefont {P.~J.}\ \bibnamefont {Mulders}}, \ and\ \bibinfo {author}
  {\bibfnamefont {M.}~\bibnamefont {Schlegel}},\ }\href
  {http://dx.doi.org/10.1088/1126-6708/2007/02/093} {\bibinfo  {journal} {J.
  High Energy Phys. 02 (2007) 093}\ }\BibitemShut {NoStop}%
\bibitem [{\citenamefont {Barone}\ \emph {et~al.}(2010)\citenamefont {Barone},
  \citenamefont {Bradamante},\ and\ \citenamefont
  {Martin}}]{Barone_review_2010}%
  \BibitemOpen
\bibfield  {journal} {  }\bibfield  {author} {\bibinfo {author} {\bibfnamefont
  {V.}~\bibnamefont {Barone}}, \bibinfo {author} {\bibfnamefont
  {F.}~\bibnamefont {Bradamante}}, \ and\ \bibinfo {author} {\bibfnamefont
  {A.}~\bibnamefont {Martin}},\ }\href {\doibase 10.1016/j.ppnp.2010.07.003}
  {\bibfield  {journal} {\bibinfo  {journal} {Prog. Part. Nucl. Phys.}\
  }\textbf {\bibinfo {volume} {65}},\ \bibinfo {pages} {267} (\bibinfo {year}
  {2010})}\BibitemShut {NoStop}%
\bibitem [{\citenamefont {Barone}\ \emph {et~al.}(2015)\citenamefont {Barone}
  \emph {et~al.}}]{Barone2015}%
  \BibitemOpen
  \bibfield  {author} {\bibinfo {author} {\bibfnamefont {V.}~\bibnamefont
  {Barone}} \emph {et~al.},\ }\href {\doibase 10.1103/PhysRevD.91.074019}
  {\bibfield  {journal} {\bibinfo  {journal} {Phys. Rev. D}\ }\textbf {\bibinfo
  {volume} {91}},\ \bibinfo {pages} {074019} (\bibinfo {year}
  {2015})}\BibitemShut {NoStop}%
\bibitem [{\citenamefont {Anselmino}\ \emph {et~al.}()\citenamefont
  {Anselmino}, \citenamefont {Boglione}, \citenamefont {Gonzalez~Hernandez},
  \citenamefont {Melis},\ and\ \citenamefont {Prokudin}}]{Anselmino2014}%
  \BibitemOpen
  \bibfield  {author} {\bibinfo {author} {\bibfnamefont {M.}~\bibnamefont
  {Anselmino}}, \bibinfo {author} {\bibfnamefont {M.}~\bibnamefont {Boglione}},
  \bibinfo {author} {\bibfnamefont {J.~O.}\ \bibnamefont {Gonzalez~Hernandez}},
  \bibinfo {author} {\bibfnamefont {S.}~\bibnamefont {Melis}}, \ and\ \bibinfo
  {author} {\bibfnamefont {A.}~\bibnamefont {Prokudin}},\ }\href
  {http://dx.doi.org/10.1007/JHEP04(2014)005} {\bibinfo  {journal} {J. High
  Energy Phys. 04 (2014) 005}\ }\BibitemShut {NoStop}%
\bibitem [{\citenamefont {Bacchetta}\ and\ \citenamefont
  {Radici}(2011)}]{Bacchetta2011}%
  \BibitemOpen
\bibfield  {journal} {  }\bibfield  {author} {\bibinfo {author} {\bibfnamefont
  {A.}~\bibnamefont {Bacchetta}}\ and\ \bibinfo {author} {\bibfnamefont
  {M.}~\bibnamefont {Radici}},\ }\href {\doibase
  10.1103/PhysRevLett.107.212001} {\bibfield  {journal} {\bibinfo  {journal}
  {Phys. Rev. Lett.}\ }\textbf {\bibinfo {volume} {107}},\ \bibinfo {pages}
  {212001} (\bibinfo {year} {2011})}\BibitemShut {NoStop}%
\bibitem [{\citenamefont {Cahn}(1978)}]{Cahn_original}%
  \BibitemOpen
  \bibfield  {author} {\bibinfo {author} {\bibfnamefont {R.~N.}\ \bibnamefont
  {Cahn}},\ }\href {\doibase 10.1016/0370-2693(78)90020-5} {\bibfield
  {journal} {\bibinfo  {journal} {Phys. Lett. B}\ }\textbf {\bibinfo {volume}
  {78}},\ \bibinfo {pages} {269} (\bibinfo {year} {1978})}\BibitemShut
  {NoStop}%
\bibitem [{\citenamefont {Collins}(2013)}]{Collins_book}%
  \BibitemOpen
  \bibfield  {author} {\bibinfo {author} {\bibfnamefont {J.}~\bibnamefont
  {Collins}},\ }\href {http://www.cambridge.org/de/knowledge/isbn/item5756723}
  {\emph {\bibinfo {title} {{Foundations of perturbative QCD}}}}\ (\bibinfo
  {publisher} {Cambridge University Press},\ \bibinfo {year}
  {2013})\BibitemShut {NoStop}%
\bibitem [{\citenamefont {Olive}\ \emph {et~al.}(2014)\citenamefont {Olive}
  \emph {et~al.}}]{PDG}%
  \BibitemOpen
  \bibfield  {author} {\bibinfo {author} {\bibfnamefont {K.~A.}\ \bibnamefont
  {Olive}} \emph {et~al.} (\bibinfo {collaboration} {Particle Data Group}),\
  }\href {\doibase 10.1088/1674-1137/38/9/090001} {\bibfield  {journal}
  {\bibinfo  {journal} {Chin. Phys. C}\ }\textbf {\bibinfo {volume} {38}},\
  \bibinfo {pages} {090001} (\bibinfo {year} {2014})}\BibitemShut {NoStop}%
\bibitem [{\citenamefont {Kuhn}\ \emph {et~al.}(2009)\citenamefont {Kuhn},
  \citenamefont {Chen},\ and\ \citenamefont {Leader}}]{crisis}%
  \BibitemOpen
  \bibfield  {author} {\bibinfo {author} {\bibfnamefont {S.}~\bibnamefont
  {Kuhn}}, \bibinfo {author} {\bibfnamefont {J.-P.}\ \bibnamefont {Chen}}, \
  and\ \bibinfo {author} {\bibfnamefont {E.}~\bibnamefont {Leader}},\ }\href
  {\doibase http://dx.doi.org/10.1016/j.ppnp.2009.02.001} {\bibfield  {journal}
  {\bibinfo  {journal} {Prog. in Part. and Nucl. Phys.}\ }\textbf {\bibinfo
  {volume} {63}},\ \bibinfo {pages} {1 } (\bibinfo {year} {2009})}\BibitemShut
  {NoStop}%
\bibitem [{\citenamefont {Asaturyan}\ \emph {et~al.}(2012)\citenamefont
  {Asaturyan} \emph {et~al.}}]{HallC_PRC}%
  \BibitemOpen
  \bibfield  {author} {\bibinfo {author} {\bibfnamefont {R.}~\bibnamefont
  {Asaturyan}} \emph {et~al.},\ }\href {\doibase 10.1103/PhysRevC.85.015202}
  {\bibfield  {journal} {\bibinfo  {journal} {Phys. Rev. C}\ }\textbf {\bibinfo
  {volume} {85}},\ \bibinfo {pages} {015202} (\bibinfo {year}
  {2012})}\BibitemShut {NoStop}%
\bibitem [{\citenamefont {Osipenko}\ \emph {et~al.}(2009)\citenamefont
  {Osipenko} \emph {et~al.}}]{CLAS_PRD}%
  \BibitemOpen
  \bibfield  {author} {\bibinfo {author} {\bibfnamefont {M.}~\bibnamefont
  {Osipenko}} \emph {et~al.} (\bibinfo {collaboration} {CLAS}),\ }\href
  {\doibase 10.1103/PhysRevD.80.032004} {\bibfield  {journal} {\bibinfo
  {journal} {Phys. Rev. D}\ }\textbf {\bibinfo {volume} {80}},\ \bibinfo
  {pages} {032004} (\bibinfo {year} {2009})}\BibitemShut {NoStop}%
\bibitem [{\citenamefont {Airapetian}\ \emph
  {et~al.}(2013{\natexlab{a}})\citenamefont {Airapetian} \emph
  {et~al.}}]{HERMES_1}%
  \BibitemOpen
  \bibfield  {author} {\bibinfo {author} {\bibfnamefont {A.}~\bibnamefont
  {Airapetian}} \emph {et~al.} (\bibinfo {collaboration} {HERMES
  Collaboration}),\ }\href {\doibase 10.1103/PhysRevD.87.074029} {\bibfield
  {journal} {\bibinfo  {journal} {Phys. Rev. D}\ }\textbf {\bibinfo {volume}
  {87}},\ \bibinfo {pages} {074029} (\bibinfo {year}
  {2013}{\natexlab{a}})}\BibitemShut {NoStop}%
\bibitem [{\citenamefont {Adolph}\ \emph {et~al.}(2013)\citenamefont {Adolph}
  \emph {et~al.}}]{COMPASS_1}%
  \BibitemOpen
  \bibfield  {author} {\bibinfo {author} {\bibfnamefont {C.}~\bibnamefont
  {Adolph}} \emph {et~al.} (\bibinfo {collaboration} {COMPASS}),\ }\href
  {\doibase 10.1140/epjc/s10052-013-2531-6, 10.1140/epjc/s10052-014-3255-y}
  {\bibfield  {journal} {\bibinfo  {journal} {Eur. Phys. J. C}\ }\textbf
  {\bibinfo {volume} {73}},\ \bibinfo {pages} {2531} (\bibinfo {year}
  {2013})}\BibitemShut {NoStop}%
\bibitem [{\citenamefont {Airapetian}\ \emph
  {et~al.}(2013{\natexlab{b}})\citenamefont {Airapetian} \emph
  {et~al.}}]{HERMES_2}%
  \BibitemOpen
  \bibfield  {author} {\bibinfo {author} {\bibfnamefont {A.}~\bibnamefont
  {Airapetian}} \emph {et~al.} (\bibinfo {collaboration} {HERMES
  Collaboration}),\ }\href {\doibase 10.1103/PhysRevD.87.012010} {\bibfield
  {journal} {\bibinfo  {journal} {Phys. Rev. D}\ }\textbf {\bibinfo {volume}
  {87}},\ \bibinfo {pages} {012010} (\bibinfo {year}
  {2013}{\natexlab{b}})}\BibitemShut {NoStop}%
\bibitem [{\citenamefont {Adolph}\ \emph {et~al.}(2014)\citenamefont {Adolph}
  \emph {et~al.}}]{COMPASS_2}%
  \BibitemOpen
  \bibfield  {author} {\bibinfo {author} {\bibfnamefont {C.}~\bibnamefont
  {Adolph}} \emph {et~al.} (\bibinfo {collaboration} {COMPASS}),\ }\href
  {\doibase 10.1016/j.nuclphysb.2014.07.019} {\bibfield  {journal} {\bibinfo
  {journal} {Nucl. Phys. B}\ }\textbf {\bibinfo {volume} {886}},\ \bibinfo
  {pages} {1046} (\bibinfo {year} {2014})}\BibitemShut {NoStop}%
\bibitem [{\citenamefont {Friar}\ \emph {et~al.}(1990)\citenamefont {Friar},
  \citenamefont {Gibson}, \citenamefont {Payne}, \citenamefont {Bernstein},\
  and\ \citenamefont {Chupp}}]{3He_spin}%
  \BibitemOpen
  \bibfield  {author} {\bibinfo {author} {\bibfnamefont {J.~L.}\ \bibnamefont
  {Friar}}, \bibinfo {author} {\bibfnamefont {B.~F.}\ \bibnamefont {Gibson}},
  \bibinfo {author} {\bibfnamefont {G.~L.}\ \bibnamefont {Payne}}, \bibinfo
  {author} {\bibfnamefont {A.~M.}\ \bibnamefont {Bernstein}}, \ and\ \bibinfo
  {author} {\bibfnamefont {T.~E.}\ \bibnamefont {Chupp}},\ }\href {\doibase
  10.1103/PhysRevC.42.2310} {\bibfield  {journal} {\bibinfo  {journal} {Phys.
  Rev. C}\ }\textbf {\bibinfo {volume} {42}},\ \bibinfo {pages} {2310}
  (\bibinfo {year} {1990})}\BibitemShut {NoStop}%
\bibitem [{\citenamefont {Qian}\ \emph {et~al.}(2011)\citenamefont {Qian} \emph
  {et~al.}}]{XQ}%
  \BibitemOpen
  \bibfield  {author} {\bibinfo {author} {\bibfnamefont {X.}~\bibnamefont
  {Qian}} \emph {et~al.},\ }\href {\doibase 10.1103/PhysRevLett.107.072003}
  {\bibfield  {journal} {\bibinfo  {journal} {Phys. Rev. Lett.}\ }\textbf
  {\bibinfo {volume} {107}},\ \bibinfo {pages} {072003} (\bibinfo {year}
  {2011})}\BibitemShut {NoStop}%
\bibitem [{\citenamefont {Huang}\ \emph {et~al.}(2012)\citenamefont {Huang}
  \emph {et~al.}}]{JH}%
  \BibitemOpen
  \bibfield  {author} {\bibinfo {author} {\bibfnamefont {J.}~\bibnamefont
  {Huang}} \emph {et~al.},\ }\href {\doibase 10.1103/PhysRevLett.108.052001}
  {\bibfield  {journal} {\bibinfo  {journal} {Phys. Rev. Lett.}\ }\textbf
  {\bibinfo {volume} {108}},\ \bibinfo {pages} {052001} (\bibinfo {year}
  {2012})}\BibitemShut {NoStop}%
\bibitem [{\citenamefont {Zhang}\ \emph {et~al.}(2014)\citenamefont {Zhang}
  \emph {et~al.}}]{YZ}%
  \BibitemOpen
  \bibfield  {author} {\bibinfo {author} {\bibfnamefont {Y.}~\bibnamefont
  {Zhang}} \emph {et~al.},\ }\href {\doibase 10.1103/PhysRevC.90.055209}
  {\bibfield  {journal} {\bibinfo  {journal} {Phys. Rev. C}\ }\textbf {\bibinfo
  {volume} {90}},\ \bibinfo {pages} {055209} (\bibinfo {year}
  {2014})}\BibitemShut {NoStop}%
\bibitem [{\citenamefont {Zhao}\ \emph {et~al.}(2014)\citenamefont {Zhao} \emph
  {et~al.}}]{YXZ}%
  \BibitemOpen
  \bibfield  {author} {\bibinfo {author} {\bibfnamefont {Y.~X.}\ \bibnamefont
  {Zhao}} \emph {et~al.},\ }\href {\doibase 10.1103/PhysRevC.90.055201}
  {\bibfield  {journal} {\bibinfo  {journal} {Phys. Rev. C}\ }\textbf {\bibinfo
  {volume} {90}},\ \bibinfo {pages} {055201} (\bibinfo {year}
  {2014})}\BibitemShut {NoStop}%
\bibitem [{\citenamefont {Anselmino}(2016)}]{Anselmino_phm}%
  \BibitemOpen
  \bibfield  {author} {\bibinfo {author} {\bibfnamefont {M.}~\bibnamefont
  {Anselmino}},\ }\bibfield  {booktitle} {\emph {\bibinfo {booktitle}
  {{Proceedings, Theory and Experiment for Hadrons on the Light-Front (Light
  Cone 2015)}}},\ }\href {\doibase 10.1007/s00601-016-1072-6} {\bibfield
  {journal} {\bibinfo  {journal} {Few Body Syst.}\ }\textbf {\bibinfo {volume}
  {57}},\ \bibinfo {pages} {373} (\bibinfo {year} {2016})}\BibitemShut
  {NoStop}%
\bibitem [{\citenamefont {Martin}(2016)}]{Martin_2016}%
  \BibitemOpen
  \bibfield  {author} {\bibinfo {author} {\bibfnamefont {A.}~\bibnamefont
  {Martin}},\ }\bibfield  {booktitle} {\emph {\bibinfo {booktitle}
  {{Proceedings, 21st International Symposium on Spin Physics (SPIN 2014)}}},\
  }\href {\doibase 10.1142/S2010194516600284} {\bibfield  {journal} {\bibinfo
  {journal} {Int. J. Mod. Phys. Conf. Ser.}\ }\textbf {\bibinfo {volume}
  {40}},\ \bibinfo {pages} {1660028} (\bibinfo {year} {2016})}\BibitemShut
  {NoStop}%
\bibitem [{\citenamefont {Bacchetta}\ \emph {et~al.}(2004)\citenamefont
  {Bacchetta}, \citenamefont {D'Alesio}, \citenamefont {Diehl},\ and\
  \citenamefont {Miller}}]{trento_convention}%
  \BibitemOpen
  \bibfield  {author} {\bibinfo {author} {\bibfnamefont {A.}~\bibnamefont
  {Bacchetta}}, \bibinfo {author} {\bibfnamefont {U.}~\bibnamefont {D'Alesio}},
  \bibinfo {author} {\bibfnamefont {M.}~\bibnamefont {Diehl}}, \ and\ \bibinfo
  {author} {\bibfnamefont {C.~A.}\ \bibnamefont {Miller}},\ }\href {\doibase
  10.1103/PhysRevD.70.117504} {\bibfield  {journal} {\bibinfo  {journal} {Phys.
  Rev. D}\ }\textbf {\bibinfo {volume} {70}},\ \bibinfo {pages} {117504}
  (\bibinfo {year} {2004})}\BibitemShut {NoStop}%
\bibitem [{\citenamefont {Signori}\ \emph {et~al.}()\citenamefont {Signori},
  \citenamefont {Bacchetta}, \citenamefont {Radici},\ and\ \citenamefont
  {Schnell}}]{TMD_width_flavor}%
  \BibitemOpen
  \bibfield  {author} {\bibinfo {author} {\bibfnamefont {A.}~\bibnamefont
  {Signori}}, \bibinfo {author} {\bibfnamefont {A.}~\bibnamefont {Bacchetta}},
  \bibinfo {author} {\bibfnamefont {M.}~\bibnamefont {Radici}}, \ and\ \bibinfo
  {author} {\bibfnamefont {G.}~\bibnamefont {Schnell}},\ }\href
  {http://dx.doi.org/10.1007/JHEP11(2013)194} {\bibinfo  {journal} {J. High
  Energy Phys. 11 (2013) 194}\ }\BibitemShut {NoStop}%
\bibitem [{\citenamefont {HAPPER}(1972)}]{seop}%
  \BibitemOpen
\bibfield  {journal} {  }\bibfield  {author} {\bibinfo {author} {\bibfnamefont
  {W.}~\bibnamefont {HAPPER}},\ }\href {\doibase 10.1103/RevModPhys.44.169}
  {\bibfield  {journal} {\bibinfo  {journal} {Rev. Mod. Phys.}\ }\textbf
  {\bibinfo {volume} {44}},\ \bibinfo {pages} {169} (\bibinfo {year}
  {1972})}\BibitemShut {NoStop}%
\bibitem [{\citenamefont {Qian}(2011)}]{xq_thesis}%
  \BibitemOpen
  \bibfield  {author} {\bibinfo {author} {\bibfnamefont {X.}~\bibnamefont
  {Qian}},\ }\href {http://hallaweb.jlab.org/experiment/transversity/thesis}
  {Ph.D. thesis},\ \bibinfo  {school} {Duke} (\bibinfo {year}
  {2011})\BibitemShut {NoStop}%
\bibitem [{\citenamefont {Allada}\ \emph {et~al.}(2014)\citenamefont {Allada}
  \emph {et~al.}}]{KA}%
  \BibitemOpen
  \bibfield  {author} {\bibinfo {author} {\bibfnamefont {K.}~\bibnamefont
  {Allada}} \emph {et~al.},\ }\href {\doibase 10.1103/PhysRevC.89.042201}
  {\bibfield  {journal} {\bibinfo  {journal} {Phys. Rev. C}\ }\textbf {\bibinfo
  {volume} {89}},\ \bibinfo {pages} {042201} (\bibinfo {year}
  {2014})}\BibitemShut {NoStop}%
\bibitem [{\citenamefont {Alcorn}\ \emph {et~al.}(2004)\citenamefont {Alcorn}
  \emph {et~al.}}]{HRS_NIM}%
  \BibitemOpen
  \bibfield  {author} {\bibinfo {author} {\bibfnamefont {J.}~\bibnamefont
  {Alcorn}} \emph {et~al.},\ }\href {\doibase 10.1016/j.nima.2003.11.415}
  {\bibfield  {journal} {\bibinfo  {journal} {Nucl. Instrum. Meth.}\ }\textbf
  {\bibinfo {volume} {A522}},\ \bibinfo {pages} {294} (\bibinfo {year}
  {2004})}\BibitemShut {NoStop}%
\bibitem [{SIM({\natexlab{a}})}]{SIMC_pack}%
  \BibitemOpen
  \href {https://hallcweb.jlab.org/wiki/index.php/Monte_Carlo} {\emph {\bibinfo
  {title} {{Hall C Monte Carlo package for SIDIS}}}}\BibitemShut {NoStop}%
\bibitem [{SIM({\natexlab{b}})}]{SIMC_trans}%
  \BibitemOpen
  \href
  {https://userweb.jlab.org/~puckett/e06010/simc_transversity_documentation.pd%
f} {\emph {\bibinfo {title} {{SIMC for E06-010}}}}\BibitemShut {NoStop}%
\bibitem [{\citenamefont {Akushevich}\ \emph {et~al.}(2009)\citenamefont
  {Akushevich}, \citenamefont {Ilyichev},\ and\ \citenamefont
  {Osipenko}}]{HAPRAD}%
  \BibitemOpen
  \bibfield  {author} {\bibinfo {author} {\bibfnamefont {I.}~\bibnamefont
  {Akushevich}}, \bibinfo {author} {\bibfnamefont {A.}~\bibnamefont
  {Ilyichev}}, \ and\ \bibinfo {author} {\bibfnamefont {M.}~\bibnamefont
  {Osipenko}},\ }\href {\doibase 10.1016/j.physletb.2008.12.058} {\bibfield
  {journal} {\bibinfo  {journal} {Phys. Lett. B}\ }\textbf {\bibinfo {volume}
  {672}},\ \bibinfo {pages} {35} (\bibinfo {year} {2009})}\BibitemShut
  {NoStop}%
\bibitem [{\citenamefont {Mo}\ and\ \citenamefont {Tsai}(1969)}]{MoTsai}%
  \BibitemOpen
  \bibfield  {author} {\bibinfo {author} {\bibfnamefont {L.~W.}\ \bibnamefont
  {Mo}}\ and\ \bibinfo {author} {\bibfnamefont {Y.-S.}\ \bibnamefont {Tsai}},\
  }\href {\doibase 10.1103/RevModPhys.41.205} {\bibfield  {journal} {\bibinfo
  {journal} {Rev. Mod. Phys.}\ }\textbf {\bibinfo {volume} {41}},\ \bibinfo
  {pages} {205} (\bibinfo {year} {1969})}\BibitemShut {NoStop}%
\bibitem [{\citenamefont {Gao}\ \emph {et~al.}(2011)\citenamefont {Gao} \emph
  {et~al.}}]{Jlab12GeV}%
  \BibitemOpen
  \bibfield  {author} {\bibinfo {author} {\bibfnamefont {H.}~\bibnamefont
  {Gao}} \emph {et~al.},\ }\href {\doibase 10.1140/epjp/i2011-11002-4}
  {\bibfield  {journal} {\bibinfo  {journal} {Eur. Phys. J.}\ }\textbf
  {\bibinfo {volume} {126}},\ \bibinfo {pages} {1} (\bibinfo {year}
  {2011})}\BibitemShut {NoStop}%
\bibitem [{\citenamefont {Chen}\ \emph {et~al.}()\citenamefont {Chen} \emph
  {et~al.}}]{SoLID_white_paper}%
  \BibitemOpen
  \bibfield  {author} {\bibinfo {author} {\bibfnamefont {J.-P.}\ \bibnamefont
  {Chen}} \emph {et~al.},\ }\href {http://arxiv.org/abs/1409.7741} {\bibinfo
  {journal} {arXiv:1409.7741}\ }\BibitemShut {NoStop}%
\end{thebibliography}%

\end{document}